# Lateral Contrast Enhancement in Tomographic Volumetric 3D-Printing via Binary Photoinhibition

Bin Wang[1], Weichao Sun[2,3], Hossein Safari[2], S. Kaveh Hedayati[1], Jadze P. C. Narag[4], Thor D. V. Christiansen[5], Adrian A. Schiefler[6], Henning O. Sørensen[6,7], Jeppe R. Frisvad[5], Kristoffer Almdal[2], Aminul Islam[1], Yi Yang[2,8]*

[1] Department of Civil and Mechanical Engineering, Technical University of Denmark; 2800 Kongens Lyngby, Denmark.

[2] Department of Chemistry, Technical University of Denmark; 2800 Kongens Lyngby, Denmark.

[3] PERFI Technologies ApS; 2800 Kongens Lyngby, Denmark.

[4] Department of Health Technology, Technical University of Denmark; 2800 Kongens Lyngby, Denmark.

[5] Department of Applied Mathematics and Computer Science, Technical University of Denmark; 2800 Kongens Lyngby, Denmark.

[6] Department of Physics, Technical University of Denmark; 2800 Kongens Lyngby, Denmark.

[7] Xnovo Technology ApS, 4600 Køge, Denmark.

[8] Center for Energy Resources Engineering, Technical University of Denmark; 2800 Kongens Lyngby, Denmark.

* To whom correspondence should be addressed (email: yyan@dtu.dk)

**Abstract**

Tomographic volumetric 3D-printing (TVP) utilizes a nonlinear photoresponse of polymer precursor to cure all points in a three-dimensional (3D) object in parallel. A key challenge in TVP is to build up dose contrast between in-part and out-of-part points in a lateral plane, which relies on coordinated illumination from various projecting angles. This challenge has mainly been tackled by projection optimization. Here we show that designing material responses to photo-excitation can be a more effective way of addressing this challenge. By introducing a secondary photo-inhibitory species that reacts to external ultraviolet (UV) stimulus, we create a binary photoinhibition (BPI) system that greatly enhances the achievable dose contrast in a lateral plane. We first show that, in theory, combining dose subtraction with sufficient projection angles can guarantee an exact mathematical reconstruction of any greyscale design. We then propose a theoretical framework for BPI, in which a single stationary state with swappable stability can be used to realize dose subtraction. We use oxygen-lophyl radical pair as an approximation to show improvements in print quality enabled by enhanced dose contrast. *In situ* shadowgraphy shows that BPI improves the lateral patterning with various geometric features, creating differentiable changes in refractive index within 54 μm or less. We show qualitative improvements in surface features and internal hollowness in physical prints. The direct impacts of UV light on the formation of positive and negative features on vertical and lateral planes of 5 workpieces are analyzed quantitatively. We conclude that introducing BPI with UV irradiation grants us direct control over the formation of negative features on the lateral plane. The results presents new opportunities in addressing a long-standing challenge in TVP development, and offers new insights into the printability of photoresins using TVP.

**Introduction**

Tomographic volumetric 3D-printing (TVP) is an emerging vat photopolymerization method that enables high-speed fabrication of objects with complex geometry. Pioneering works in this area[1, 2, 3, 4] utilize a nonlinear photoresponse of polymer precursors to solidify all points in a 3D-object in parallel. As a result, the method severs the inherent link between the geometric complexity of a workpiece and its fabrication time. This "volume-at-once" advantage of TVP makes it a promising volumetric additive manufacturing (VAM) approach for building structures that require a massive number of voxels to describe. In addition to printing speed, TVP also offers perquisites such as smooth surface finish, minimum auxiliary support, and the ability to 3D-print over existing objects. However, the reliance on photopolymerization entails restrictions for chemical, optical and mechanical properties of photoresins as well as the patterned projections delivering light doses from various angles. The majority of validated polymer precursors rely on selectively inhibited free radical polymerization to create a desired nonlinear photoresponse. For acrylate/methacrylate-based precursors, this can be done by utilizing pre-dissolved oxygen from the atmosphere,[1, 2, 3, 5, 6, 7, 8, 9, 10, 11, 12, 13, 14, 15, 16, 17, 18] or by including 2,2,6,6-tetramethyl-1-piperidinyloxy (TEMPO) in the resin formulation[14]. TEMPO has also been used to introduce induction periods for thiol-ene resins,[4] acrylate-epoxy mixtures[19] and silica nanocomposites[16, 20]. The concentrations of photo-initiators and inhibitors are tweaked to strike a balance between attenuation and absorption across the curing volume. TVP has been used to process hydrogel, thermoplastic, thermoset, silica nanocomposite and preceramic precursors: Table 1 gives a non-exhaustive summary of published validations. For comparison, we have also included two seminal works on a light-sheet based method that requires dual-color photoinitiation.[21, 22]

One of the key traits of TVP is its anisotropy. The smallest features producible using TVP depends on the orientation of the workpiece, due to the different building mechanisms of dose contrast in the vertical and lateral planes. Here, "dose contrast" is the difference in the dose of irradiation received by two points in the curing space. Assuming a classic parallel beam setup, the vertical plane is the plane normal to the direction of light propagation (parallel to the focal plane of a projector). The dose contrast within this plane can be created directly by patterning the projection, and there is a clear correspondence between the smallest features producible in the vertical plane and the projector's optical characteristics (e.g., pixel pitch on the focal plane). Conversely, the lateral plane is parallel to the direction of light projection and perpendicular to the axis of rotation. The contrast in this plane can only be built indirectly by coordinating projections from multiple angles. Because each ray illuminates an entire line, out-of-part points (points in the curing volume that are not supposed to solidify) would inevitably receive doses for in-part points that fall on the same lines (see, for example in Fig.1a-b, to deliver dose of illumination to point $(x, y)$, all points on the line section $l$ will receive irradiation). If we define the cumulative distribution of received irradiation in the curing volume as a dose "buildup", it is physically impossible to subtract excessive doses from an existing buildup. Hence, it is difficult to avoid over-exposure that leads to unintended polymerization in a lateral plane. Many researchers have emphasized the mechanistic differences between vertical and lateral planes in TVP,[3, 23, 24] as well as in producing positive and negative features within the same plane.[9, 25] However, it remains elusive whether the smallest feature observed in a printed object can be claimed as TVP's lateral resolution, because some features can only be created in simpler designs and the same feature size may not be reproducible in a more complex lateral pattern. In addition, it remains unclear whether a relation between lateral print resolution and the optical characteristics of the projector used can be defined. In summary, a key challenge for TVP is to maximize the dose contrast within the lateral plane, and ideally the geometric design (the target pattern) should have minimum impact on this achievable contrast.

To the best of our knowledge, apart from prolonging induction by adding TEMPO (discussed below as a printability issue), contrast building on the lateral plane for TVP has mainly been enhanced via optimization of projecting patterns. For example, Bhattacharya et al. introduced dose matching in 2019[1] and penalty minimization in 2021.[6] Rackson et al. introduced object-space model optimization in 2021,[26] which has later been adopted in a series of works.[13, 14, 15, 20] The three methods have recently been unified into a more elegant scheme called band-constraint $Lp$ norm minimization.[27] Chen et al. presented iterative photomask optimization via expectation maximization in a recent work.[17] In addition, ray tracing has been used to account for non-ideal behaviors of light propagation, such as non-telecentricity, local scattering, refraction and/or attenuation.[15] Diffusion and optical blurring are shown to be pre-compensable in projection calculation via deconvolution.[12] In this work, however, we present a new strategy to enhance contrast building on a lateral plane for TVP. We first present the rationale for the necessity of dose subtraction from an existing buildup using an original method of sinogram generation. We then outline a theoretical framework for creating a binary photoinhibition (BPI) system containing multiple stationary states with controllable stability. The use of BPI in TVP enables the subtraction of excessive dose from an existing buildup. We then discuss the practical constraints in its realization and demonstrate the improvements in printability using an exemplary implementation of approximated BPI. We show quantitatively the impact of a secondary writing wavelength on the formation of various features, especially on the lateral plane. We wrap up with a discussion on the deviation of the accessible system from an ideal BPI implementation, and highlight the resulting caveats.

**Table 1.** Summary of smallest features achieved in tomographic volumetric 3D-printing (TVP)

| Pixel Pitch* | Lateral Dimension | Smallest Features | Orientation of Key Features** | Characterization *** | Key Precursors | Photoinitiator {Inhibitor} | Key Wavelength | Method notes and References |
|---|---|---|---|---|---|---|---|---|
| | | 300 μm | | Photography | Bisphenol A glycerolate (1 glycerol/phenol) diacrylate Poly(ethylene glycol) diacrylate | Camphorquinone Ethyl 4-dimenthylamino benzoate | 455 nm | 1 |
| 22.8 μm | ~Ø14 mm | 145 μm (positive) 200 μm (negative) | | Microscopy | Gelatin methacrylate in phosphate buffered saline | Lithium phenyl (2,4,6-trimethylbenzoyl) phosphinate | 405 nm | 2 |
| | | | | | Triethylene glycol diallyl ether Tri-ally isocyanurate Tris[2-(3-mercaptopionyloxy)ethyl] isocyanurate | 2-methyl-4'-(methylthio)-2-morpholinopropiophenone Irgacure 907 {2,2,6,6-tetramethyl-1-piperidinyloxy} | 405 nm | 4 |
| 23 μm | ~Ø16 mm | 80 μm (positive) 500 μm (negative) | Vertical (positive) Lateral (negative) | Computed tomography | Di-pentaerythritol pentaacrylate | Phenylbis (2,4,6-trimethylbenzoyl) phosphine oxide | 405 nm | 3 |
| 65 μm | | | | | Bisphenol A glycerolate (1 glycerol/phenol) diacrylate Poly(ethylene glycol) diacrylate | Camphorquinone Ethyl 4-dimenthylamino benzoate | 460 nm | 5 |
| 31.4 μm | | | | | Urethane dimethacrylate | 2-methyl-4'-(methylthio)-2-morpholinopropiophenone | 405 nm | 6 |
| | | 200 μm | | Microscopy | Gelatin-norbornene | Lithium phenyl(2,4,6-trimethylbenzoyl) phosphinate | 405 nm | 28 Object-space model optimization[26] |
| | ~Ø16 mm | | | Photography Atomic force microscopy | Bisphenol A glycerolate (1 glycerol/phenol) diacrylate Poly(ethylene glycol) diacrylate 3,4-epoxycyclohexylmethyl 3,4-epoxycyclohexanecarboxylate | Camphorquinone Ethyl 4-dimenthylamino benzoate Triarylsulfonium hexafluoroantimonate salts | 365 nm 455 nm | Iterative forward projection[19] |
| 4.9 μm | ~4x4 mm² | 50 μm (silica) 20 μm (acrylate) | Vertical | Scanning electron microscopy | Trimethylolpropane ethoxylate triacrylate Hydroxyethylmethacrylate Silica glass nanocomposite Pentaerythritol tetraacrylate | Camphorquinone Ethyl 4-dimenthylamino benzoate Irgacure 369 {2,2,6,6-tetramethyl-1-piperidinyloxy} | 442 nm | Object-space model optimization[20] |
| 23 μm | ~Ø16 mm | 200 μm | | by estimation | Polysiloxane substituted precursor 1,4-butanediol diacrylate | Diphenyl (2,4,6-trimethylbenzoyl) phosphin oxide | 405 nm | 7 |
| | | | | | Di-pentaerythritol pentaacrylate TiO$_2$ nanoparticles Gelatin methacryloyl with cells | Phenylbis (2,4,6-trimethylbenzoyl) phosphine oxide | 405 nm | 8 |

| Pixel Pitch* | Lateral Dimension | Smallest Features | Orientation of Key Features** | Characterization *** | Key Precursors | Photoinitiator {Inhibitor} | Key Wavelength | Method notes and References |
|---|---|---|---|---|---|---|---|---|
| | | 42 μm (positive) 104 μm (negative) | Vertical | Microscopy | Gelatin methacrylate in phosphate buffered saline | Lithium phenyl (2,4,6-trimethylbenzoyl) phosphinate | 405 nm | [9] |
| 6 μm | | | | | Bisphenol A glycerolate (1 glycerol/phenol) diacrylate Poly(ethylene glycol) diacrylate Pentaerythritol tetraacrylate | Irgacure 907 Camphorquinone Ethyl 4-dimenthylamino benzoate | 405 nm 442 nm | [10] |
| | | | | | Diurethane dimethacrylate Poly(ethylene glycol) diacrylate Bisphenol A glycerolate (1 glycerol/phenol) diacrylate | Camphorquinone Ethyl 4-dimenthylamino benzoate | Blue | [29] |
| 23 μm | ~Ø30 mm | 180 μm | | Microscopy | Pentaacrylate (PRO21905, Sartomer) | Phenylbis (2,4,6-trimethylbenzoyl) phosphine oxide | 405 nm | [11] |
| 54 μm | | 195 μm | Lateral | Microscopy | Diurethane dimethacrylate Poly(ethylene glycol) diacrylate | Ethyl (2,4,6-trimethylbenzoyl) phenylphosphinate | 405 nm | Deconvolution method for diffusion and optical blurring correction[12] |
| | | 23 μm (positive) 176 μm (negative) | Vertical | Microscopy | Commercial grade gelNOR (type B, bovine hide, DoF 60%) | Lithium phenyl (2,4,6-trimethylbenzoyl) phosphinate | 405 nm | [25] |
| 14 μm | ~36x36 mm² | | | | Bisphenol A glycerolate (1 glycerol/phenol) diacrylate Poly(ethylene glycol) diacrylate | Irgacure 907 | 405 nm | Object-space model optimization[13] |
| | | 100 μm | | Scanning electron microscopy | Thiol-ene photocrosslinkable poly (ε-caprolactone) | Ethyl (2,4,6-trimethylbenzoyl) phenylphosphinate | 442 nm | [30] |
| | | | | | Norbornene and thiol-modified gelatin | Lithium phenyl (2,4,6-trimethylbenzoyl) phosphinate | 405 nm | [31] |
| 4.9 μm | ~4x4 mm² | 283 μm | | Scanning electron microscopy | Trimethylolpropane triacrylate Ethyl cellulose | Camphorquinone Ethyl 4-dimenthylamino benzoate {2,2,6,6,-tetramethyl-1-piperidinyloxy} | 442 nm | Object-space model optimization[14] |
| | | | | | Diurethane dimenthacrylate Poly(ethylene glycol) diacrylate | Camphorquinone Ethyl 4-dimenthylamino benzoate Ethyl (2,4,6-trimethylbenzoyl) phenylphosphinate | 405 nm | Ray tracing and object-space model optimization[15] |
| 50 μm | | 57 μm / 108 μm | Vertical/ Lateral | Microscopy | Silk sericin and fibroin | Ruthenium/sodium persulfate | 525 nm | [23] |

| Pixel Pitch[*] | Lateral Dimension | Smallest Features | Orientation of Key Features[**] | Characterization [***] | Key Precursors | Photoinitiator {Inhibitor} | Key Wavelength | Method notes and References |
|---|---|---|---|---|---|---|---|---|
| 23 µm | ~16x16 mm² | | | | Hydroxyethylmethacrylate<br>Trimethylolpropane ethoxylate triacrylate<br>Amorphous silica nanopowder<br>Diurethane dimethacrylate<br>Pentaerythritol tetraacrylate<br>Poly(diethoxysiloxane)<br>Triethyl borate | Diphenyl-(2,4,6-trimethylbenzoyl)-phosphinoxid<br>{2,2,6,6-tetramethyl-1-piperidinyloxy} | 405 nm | [16] |
| 50 µm | | 79 µm / 137 µm | Vertical/ Lateral | Microscopy | Hydrogel with heart-derived decellularized extracellular matrix, and meniscus-derived decellularized extracellular matrix | Tris(2,2-bipyridyl) dichlororuthenium (II) hexahydrate<br>Sodium persulfate | 525 nm | [24] |
| | | | | | Dipentaerythritol pentaacrylate<br>Poly(ethylene glycol) diacrylate<br>Dipentaerythritol pentaacrylate<br>Tricyclo [5.2.1.02,6] decanedimethanol diacrylate<br>Pentaerythritol triacrylate | Irgacure 819 | 405 nm | Iterative photomask optimization via expectation maximization[17] |
| | | | | | | | | Band-constraint $Lp$ norm minimization[27] |
| | 46.5 mm² | 250 µm | | Flatbed scanner | Trimethylolpropane triacrylate<br>Ethyl cellulose | Camphorquinone<br>Ethyl 4-dimenthylamino benzoate | 442 nm | Band-constraint $Lp$ norm minimization[18] |
| 96.6 µm | ~Ø17 mm | | | Computed tomography | Poly(ethylene glycol) diacrylate (250 Da, 575 Da, 700 Da, hydrogel) | Ethyl (2,4,6-trimethylbenzoyl) phenylphosphinate<br>Lithium phenyl-2,4,6-trimethylbenzoylphosphinate | 405 nm | Proportional-integral histogram equalization[32]<br>Root-mean-square of sign-distance functions reported |
| 21 µm | | 50 µm / 25 µm | Sweeping direction/ Focal plane | Scanning electron microscopy | Pentaerythritol tetraacrylate<br>Diurethane dimethacrylate | 5-Cyano-1,2,3,3-tetramethylindolenium iodide<br>4-Fluoro-3′-formyl-4′-hydroxybenzophenone | 375 nm<br>585 nm | Xolography actuated with a sweeping light-sheet[21] |
| 83 nm | 160x90 µm² | 2.2 µm / 500 nm | Sweeping direction/ Focal plane | Scanning electron microscopy | Dipentaerythritol hexaacrylate | 2,3-butanedione<br>{2,2,6,6-tetramethyl-1-piperidinyloxy} | 440 nm<br>660 nm | Actuated with a sweeping light-sheet[22] |

[*] on the focal plane of the projector
[**] whether the contrast required to create the feature is formed within the same lateral plane or between vertical slices
[***] metrological instrumentation based on which a printing resolution is claimed

**Necessity of dose subtraction**

In TVP, subtraction of excessive dose from an existing buildup is essential for creating dose contrast on a lateral plane. We show in this section that if dose subtraction is feasible, one could generate a sinogram that produces an exact ("perfect") greyscale dose target via direct back-projection given sufficient projecting angles. A sinogram is a mapping between physical space and projected light patterns. For simplicity, we discuss the lateral buildup of dose in a telecentric setup. Dose buildup ("**Buildup**") is defined here as the distribution of the cumulative amount of irradiation a lateral plane receives from all angles of projection according to mathematical reconstruction of a sinogram. Our goal is to create a Buildup that matches the distribution of greyscale intensity in a predefined pattern ("**Design**"). This goal can be achieved by processing points sampled from the Design sequentially to create a continuous sinogram. For example, the Design in Fig.1a has 10,000 sampling points if it is a greyscale image containing 100x100 pixels. For clarity, we refer to a sampling point in the 2D Design as a "pixel", and a point in the curing volume that corresponds to the pixel in the Design a "point". We refrained from using "voxel" in this work because the shape of a voxel in a Design (typically cubes or cuboids) can be different from that in a print (which can have fan-shaped surfaces). In a sinogram for a 360º rotation, we find a trajectory $r(\theta)$ for a point $(x, y)$ in the Buildup that corresponds to the grayscale pixel $P_1$ in the Design. Here, "trajectory" refers specifically to a curve on a sinogram that marks the relative position of the point to the source of illumination (Fig.1c shows the trajectory for $P_1$). In a telecentric setting where the origin of the sinogram ($r = 0$) is aligned with the axis of rotation (Fig.1b), we can map a point $(x, y)$ to $(\theta, r)$, where $r = y\cos\theta - x\sin\theta$ is the signed distance to the central line in the direction of illumination and $\theta$ is the angle of illumination measured from the x-axis. Let $I(\theta, r)$ be the intensity of illumination per angle that the point in question receives during a rotation, so that $D = \int_0^{2\pi} I(\theta, r) d\theta$, in which $D$ is the dose prescribed by the greyvalue of pixel $P_1$. Along the trajectory $r(\theta)$, the required intensity of illumination from the source $I'(\theta, r)$ is determined by the extinction factor ($\mu$) and the length of ray path $l$ (Fig.1d,e), i.e., $I'(\theta, r) = I(\theta, r)e^{\mu l}$, in which $l = \sqrt{R^2 - r^2} - x\cos\theta - y\sin\theta$ and $R$ is the radius of the curing volume (Fig.1b). $R$ and $l$ are measured in pixels and $\mu \cdot l$ is dimensionless. $I'(\theta, r)$ ensures that point $(x, y)$ receives the dose prescribed by the greyvalue of pixel $P_1$ in the Design after a rotation.

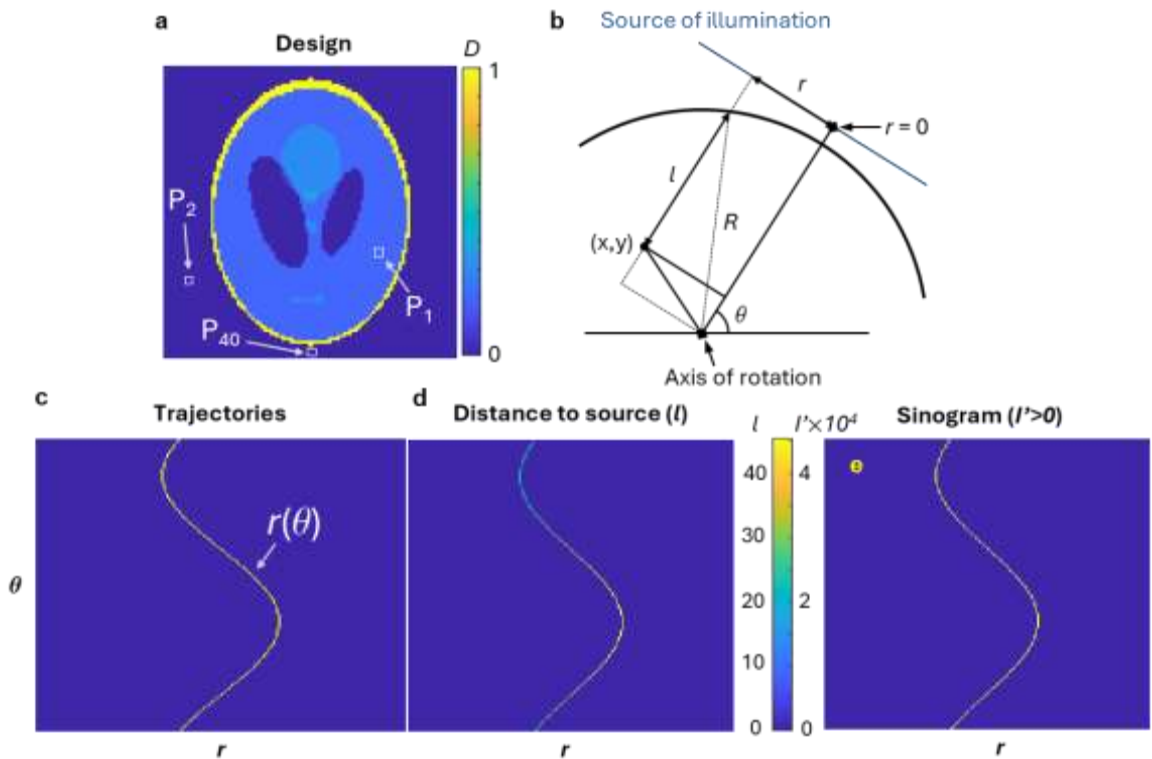

**Fig. 1**: **Sinogram for a greyscale design can be created by processing pixels sequentially.** **a** A phantom design containing 10,000 greyscale pixels ($P_1$, $P_2$, … $P_{40}$, …). **b** Schematic drawing of a telecentric setup in which the axis of rotation is aligned with the center of the projection (source of illumination). **c** The trajectory, $r(\theta)$, for $P_1$. **d** The distance between the point in the curing volume that corresponds to pixel $P_1$ in the Design and the source of illumination can be quantified along the trajectory. The color indicates dimensionless distance

measured in number of pixels when the diagonal of the Design is scaled to the diameter of the curing volume. **e** The intensity of illumination ($I'$) along the trajectory when $\mu \cdot R = 1$. The color indicates dimensionless intensity measured in the greyvalue of $P_1$. In **e** the doses received by the point in question from all projecting angles are identical.

Moving onward, for the *n*-th pixel in the Design, its trajectory $r_n(\theta)$ in the sinogram will have two intersections with each and every trajectory of the *n*-1 pixels that had been processed, corresponding to the two projection angles along which the pixel in question falls in line with a processed pixel. The unintended dose $D'_n$ – the mandatory amount of irradiation that the *n*-th pixel receives at these intersections, is given by $D'_n = \sum_{i=1}^{2(n-1)} I'(\theta_i, y_n \cos\theta_i - x_n \sin\theta_i) e^{-\mu l_n} \Delta\theta$. Let $\Delta D_n = D_n - D'_n$ be the mismatch between the prescribed and the unintended doses, and $I_n(\theta, r)$ be the intensity of illumination per angle that the point in question receives along the trajectory $r_n(\theta) = y_n \cos\theta - x_n \sin\theta$. The dose mismatch can be supplemented if $\Delta D = \int_0^{2\pi} I_n(\theta, r) \prod_{i=1}^{2(n-1)} (\theta - \theta_i) d\theta$, in which $\theta_i$ is an angle of intersection. The corresponding intensity required from the source of illumination is $I'(\theta, r) = I_n(\theta, r) e^{\mu l_n}$. The sinogram thus generated guarantees a Buildup that matches the Design via direct back projection. The sign of $I'(\theta, r)$ is determined by $\Delta D$, and is negative when the unintended dose exceeds the prescribed dose. For example, for $P_2$ the prescribed dose $D_2$ is zero (Fig.1a). Because $D_1 > 0$, the unintended doses that $(x_2, y_2)$ receives from the two intersections are positive (Fig.2a), and therefore $\Delta D_2 < 0$. As a result, the integral of $I'$ along the remaining sections of the trajectory must be negative to supplement this dose mismatch (Fig.2c). Figs 2d-h show the same procedure for $P_{40}$, for which $r_{40}(\theta)$ has 78 intersections with processed pixels (Fig.2d, processed trajectories are dimmed), from which $(x_{40}, y_{40})$ receives both positive (Fig.2g) and negative doses (Fig.2h). Figs 2i and 2j, respectively, show the sinograms for dose addition and subtraction after processing 40 pixels, in which significant negativities are required to eliminate $\Delta D$. Fig.2k and 2l show the sinograms generated following these steps for the entire phantom Design, and Fig.2m shows the resulting Buildup via direct back projection, in which positive and negative illuminations had been assigned different extinction factors (the MATLAB implementation of this algorithm is supplemented as Script.S1). In summary, dose subtraction ($I' < 0$) is a prerequisite to the exact reproduction of a greyscale dose target via direct back-projection. It is worth noting that no rescaling of grey value is involved in this method, and the final Buildup is a distribution of absolute (instead of rescaled) dose.

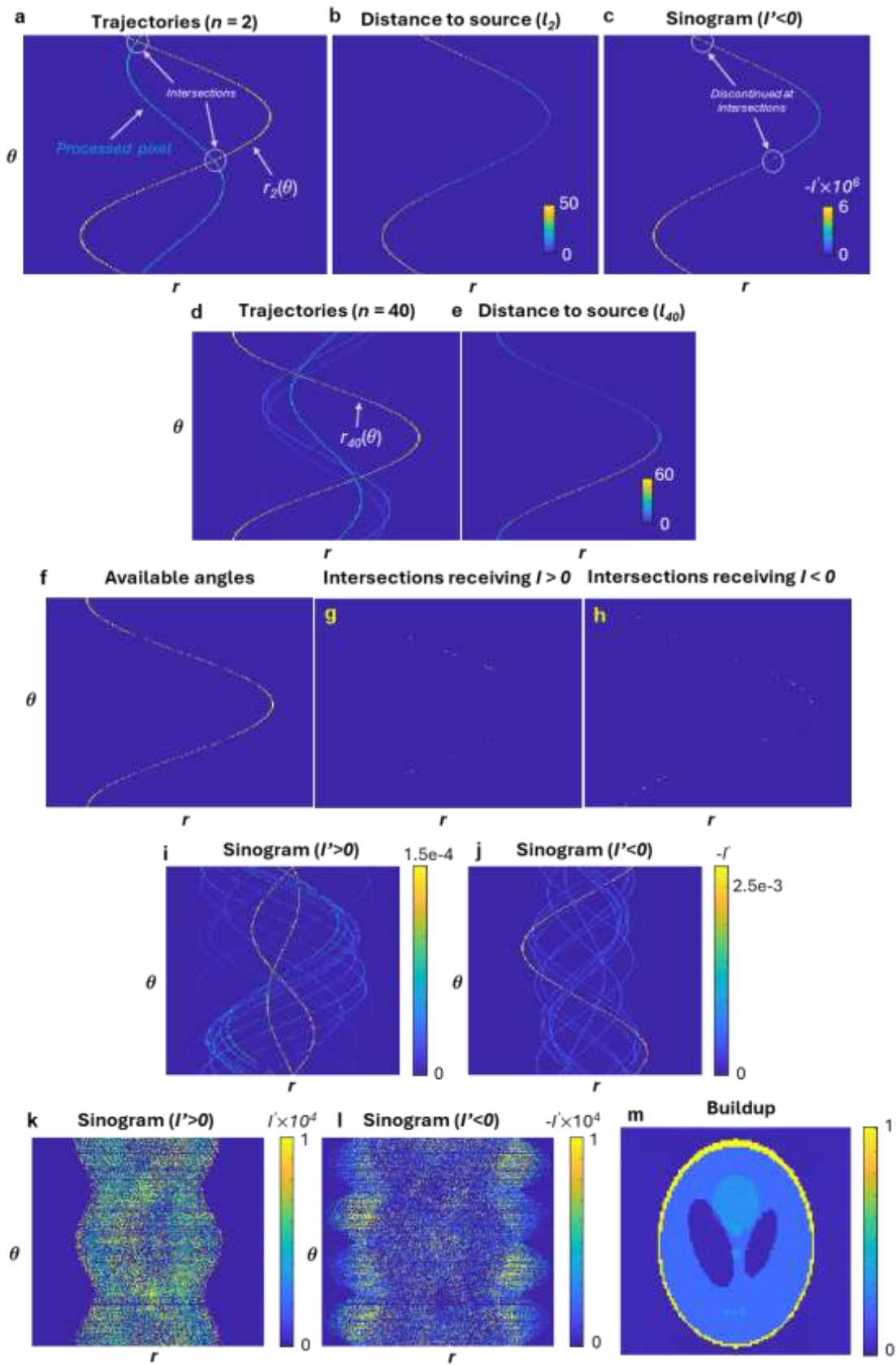
13

**Fig. 2: An exact reconstruction of a greyscale design can be realized given dose subtraction and sufficient projection angles. a** The trajectory for $P_2$ (yellow) is $r_2(\theta)$. **b** Distance between $P_2$ and the source of illumination along the trajectory, measured in number of pixels. **c** Dose subtraction is required because the point would receive more dose than prescribed by $P_2$ at the two intersections. **d** The trajectory for $P_{40}$ (yellow). The other trajectories are dimmed to emphasize $r_{40}(\theta)$. **e** Distance between $P_{40}$ and the source of illumination, measured in number of pixels. **f** Projection angles during a full rotation from which the point corresponds to $P_{40}$ does not fall on the same line as any of the previously processed 39 pixels. **g** The angles from which the point receives unintended positive doses. **h** The angles from which the point receives unintended negative doses. **i** Sinogram of positive illumination for the first 40 pixels. **j** Sinogram of negative illumination (dose subtraction) for the first 40 pixels. **k** Sinogram of positive illumination for the Design (Fig. 1a). **l** Sinogram of negative illumination for the Design (Fig. 1a). **m** Final dose Buildup via direct back-projection of the sinograms (**k, l**) without rescaling. Color scale: a, d, f, g, h – binary (blue – 0, otherwise – 1); b, e: distance measured in number of pixels; c, i, j, k, l, m: intensity measured in grey values prescribed by the Design.

**Binary photoinhibition (BPI)**

In this section, we propose a general theory that holds the potential of realizing dose subtraction in TVP. The strategy is to leverage writing light of a second wavelength to counteract the effect of the primary writing wavelength in inducing polymerization. If Species A is a pre-dissolved inhibitor to photopolymerization (e.g., oxygen or TEMPO), we look for a Species B that meets four requirements:

1. Competes with A for the incident dose of a primary wavelength that induces photopolymerization (e.g., visible light)
2. Can only be generated by irradiation of a secondary wavelength (e.g., UV light)
3. Has a long lifespan in the absence of the secondary wavelength

4. Quenches free radicals (e.g., inhibits polymerization via radical-radical termination)

With Species B, the nonlinear photoresponse that binarizes in-part and out-of-part points in single-color TVP (SC-TVP) will be replaced by a dynamic, irradiation-driven system that enables dose subtraction. The skeleton reactions and corresponding rate laws can be:

$$A \xrightarrow{Vis} A' \qquad r_1 = k_A I_{Vis} \frac{[A]}{[A]+[B]} \qquad (1)$$

$$B \xrightarrow{Vis} B' \qquad r_2 = k_B I_{Vis} \frac{[B]}{[A]+[B]} \qquad (2)$$

$$C \xrightarrow{UV} 2B \qquad r_3 = k_C I_{UV} \qquad (3)$$

$$B + B \rightarrow C \qquad r_4 = k_2 [B]^2 \qquad (4)$$

In this context, $r_i$ is the rate of elementary reaction $i$, [A] and [B] indicate the concentration (M) of Species A and B, respectively, $t$ is time (s), $k_A$, $k_B$, $k_C$ are zeroth-order rate constants (M s$^{-1}$ W$^{-1}$), $k_2$ is a second-order rate constant (M$^{-1}$s$^{-1}$), $I_{vis}$ and $I_{UV}$ are irradiance (W) of primary (visible) and secondary (UV) wavelength, respectively. The lifespan of B – the time that B remains in the absence of UV irradiation – is defined by $r_4$. The system is governed by

$$\frac{d[A]}{dt} = -k_A I_{Vis} \frac{[A]}{[A]+[B]} = F_1, \qquad (5)$$

$$\frac{d[B]}{dt} = -k_B I_{Vis} \frac{[B]}{[A]+[B]} + 2k_C I_{UV} - 2k_2 [B]^2 = F_2, \text{ and} \qquad (6)$$

$$\frac{d[C]}{dt} = -k_C I_{UV} + k_2 [B]^2 = F_3 \qquad (7)$$

On the phase diagram, only the origin represents the depletion of both inhibitors and thus the initiation of photopolymerization (Fig.3a), and a trajectory depicts how the state of a point in the curing volume evolves during printing. The trajectory always stems from the lower right corner ($[A]_0, 0$), and is steered by the irradiance $I_{vis}$ and $I_{UV}$. The illumination of the two wavelengths can be coordinated so that a system state revolves counterclockwise around the origin, i.e., $[A] \cdot F_1 + [B] \cdot F_2 = 0$. When this happens, the system evolves without getting closer or farther away from the origin, and the wavelengths effectively cancel out in terms of inducing photopolymerization: the visible light becomes photochemically "negative" to the UV light, and vice versa.

This system has a single stationary state (SS) $\left(0, \sqrt{\dfrac{2k_C I_{UV} - k_B I_{Vis}}{2k_2}}\right)$, the stability of which is given by the linearization in its neighborhood:

$$\begin{bmatrix} \dfrac{\partial F_1}{\partial [A]} & \dfrac{\partial F_1}{\partial [B]} \\ \dfrac{\partial F_2}{\partial [A]} & \dfrac{\partial F_2}{\partial [B]} \end{bmatrix}_{SS} = \begin{bmatrix} -k_A I_{Vis} \sqrt{\dfrac{2k_2}{2k_C I_{UV} - k_B I_{Vis}}} & 0 \\ k_B I_{Vis} \sqrt{\dfrac{2k_2}{2k_C I_{UV} - k_B I_{Vis}}} & -2\sqrt{2k_2 (2k_C I_{UV} - k_B I_{Vis})} \end{bmatrix}, \quad (8)$$

with eigenvalues

$$\lambda_1 \cdot \lambda_2 = 4 k_2 k_A I_{Vis}, \text{ and} \qquad (9)$$

$$\lambda_1 + \lambda_2 = -k_A I_{Vis} \sqrt{\dfrac{2k_2}{2k_C I_{UV} - k_B I_{Vis}}} - 2\sqrt{2k_2 (2k_C I_{UV} - k_B I_{Vis})}. \qquad (10)$$

The signs of eigenvalues suggest that both the position and the stability of the SS can be controlled using dual color illumination. Here, "stability" refers to the tendency of an SS recovering from small perturbations by returning to a fixed position on the phase diagram. Changing $I_{vis}$ and $I_{UV}$ can swap the SS between being in a stable state and being an oscillatory center (Figs 3b-c, Mov.S1). This controllable stability enables the pinning of out-of-part points to the upper y-axis, prolonging the period of time during which deferred termination of illumination does not result in over-exposure in TVP. Noteworthy is the similarity and difference between BPI and prior two-color VAM methods such as dual color photoinitiation (DCPI) used in xolography.[21, 22] Both methods use two colors of light to spatially confine polymerization. However, BPI is a subtractive method and the two writing wavelengths counteract each other, whereas in DCPI the two writing wavelengths are additive in triggering polymerization. BPI is designed to address the challenge in lateral reconstruction stemming from the anisotropy of TVP, and the irradiation of both wavelengths are patterned. DCPI is the cause of anisotropy of light-sheet based methods and only one of the writing light is patterned. The implementation of DCPI entails a direct connection between the print resolution and the optical characteristics of the projecting system, whereas for BPI such connection has yet to be established.

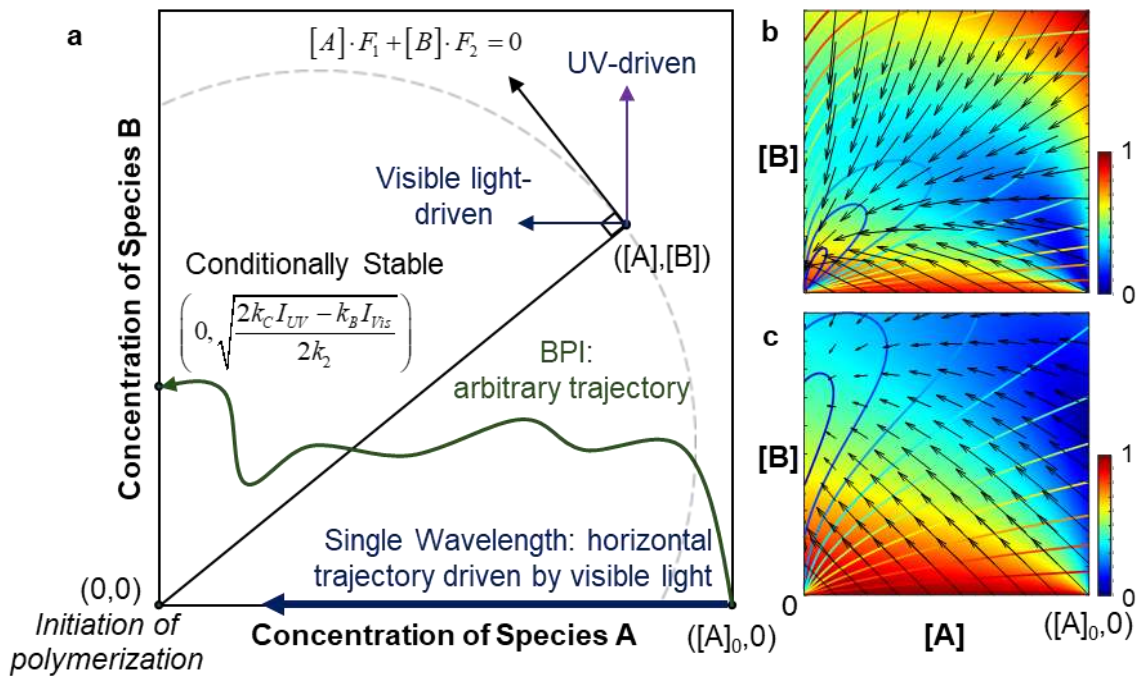

**Fig. 3: Binary photoinhibition (BPI) enables navigating the phase diagram via simultaneous use of two wavelengths. a** An illustrative phase diagram showing the advantage of having a stationary state with controllable stability. $[A]_0$ is the concentration of pre-dissolved Species A. The two wavelengths cancel out when the distance between the system state and the origin does not vary upon receiving both irradiation. **b** A desirable

diagram for an in-part point, in which the only stable stationary state is the origin, where photopolymerization initiates. **c** A desirable diagram for an out-of-part point, in which a stable stationary state is deliberately kept away from the origin using UV light. Polymerization will not occur. **b** and **c** are instant, and the color indicates relative speed of the system moving towards a stable state. See Mov.S1 for the dynamics during 3 full rotations.

Three key parameters are instrumental to understanding the effectiveness of BPI: $\alpha = k_B/k_A$, $\beta = \dfrac{k_C I_{UV}}{k_A I_{Vis}}$, and $\lambda = \ln 2/k_2$. Fig. S1 compiles the influences of these parameters. The parameter $\alpha$ marks the relative sensitivity of the two inhibitors to visible light, $\beta$ represents the relative strength of UV irradiation. Greater $\alpha$ indicates greater energy requirement in order to create an SS away from the origin. The difference between a polymerizing SS and an unpolymerized SS is more pronounced with greater $\beta$ and a Species B with a long lifespan ($\lambda$). In practice, the doses received by a point in curing volume vary periodically due to rotation, and the state of a BPI system is transient. Movie S1 shows an example of how a phase diagram evolves for a point away from the rotation axis. The efficiency of UV light can be quantified using $W = \int_0^1 \dfrac{[B]}{[A]} d\left(\dfrac{[A]}{[A]_0}\right)$. When $W = 1$, the BPI doubles the induction period. Light extinction in the curing volume, however, creates spatial variation of $W$. Fig. S2 shows the impact of the extinction factor on the achievable $W$ at five different radial positions. The scattered data were attained by tracking the phase diagram of points at fixed radial position but randomized starting angle. The closer a voxel is to the edge of the curing volume, the more sensitive its $W$ will be to the rotation periodicity. At constant UV irradiance, the effective negativity reduces towards the rotation center. Formulating a BPI can thus be demanding in that it requires a good balance between transparency and absorptivity in at least two wavelength regimes. In contrast to photoinhibitors used in layer-by-layer vat photopolymerization, Species B with a long lifespan is favored in TVP. Fig.S3 shows the impact of Species B's lifespan on achievable $W$ at three radial positions at constant UV irradiation. The energy that powers UV irradiation is spent to counteract the overspent doses of visible light that would have caused over-exposure. As a result, a dual-color setup uses more energy than its single-color counterpart for curing the same amount of material. The UV doses needed to reach $W = 1$ in Fig. S4 show the sensitivity of minimum energy requirement on radial position for producing equivalent negativity at three extinction factors.

**Practical considerations in implementing binary photoinhibition**

**Angular Resolution** With dose subtraction, the total number of points ($N$) in a Buildup that will receive the exact prescribed dose increases with the availability of projection angles. If the total number of projection angles is determined by an angular resolution $\Delta\theta$, a sufficient condition to warrant the delivery of dose mismatch $D_{N+1}$ for the $N+1$ point is $2\pi / \Delta\theta > 2N$. However, $\Delta\theta$ is lower bounded by the resolution of the projector. Consider, for example, finding a projection angle $\theta_n$ for Point C to deliver its dose mismatch $\Delta D_C$ (Fig.4). Let $\theta_i$ and $\theta_{i+1}$ be the two consecutive projection angles with the smallest angular difference at which Point C's trajectory intersects with the trajectories of all the other points, and that the two points blocking C at $\theta_i$ and $\theta_{i+1}$ are A and B. We can choose $\Delta\theta = |\theta_{i+1} - \theta_i|/2$, so that $\theta_n = \theta_i + \Delta\theta = \theta_{i+1} - \Delta\theta$. A rotation by an angular resolution $\Delta\theta$ (rotating the curing volume from $\theta + \Delta\theta$ to $\Delta\theta$) will lead to a displacement of a point's projection on the projector (Fig. 4b). The choice of $\Delta\theta$ is only meaningful if the projections of all three points on the projector ($r_A$, $r_B$ and $r_C$) fall into different pixels after such a rotation step. For any two points in the Buildup separated by distance $d$, the change of a line section $d$'s projection on the projector caused by a stepwise rotation is given by $|\Delta r' - \Delta r| = \Delta l \cdot \sin(\Delta\theta) + \Delta r \cdot [1 - \cos(\Delta\theta)]$. Here, stepwise rotation refers to the actuation of the curing volume between two consecutive projection angles spaced by the angular resolution $\Delta\theta$. Therefore, in Fig.4b, $|r_B(\theta_n) - r_C(\theta_n)| = |\Delta r'_{BC}(\theta_{i+1} - \Delta\theta) - \Delta r_{BC}(\theta_{i+1})|$ and $|r_A(\theta_n) - r_C(\theta_n)| = |\Delta r'_{AC}(\theta_i + \Delta\theta) - \Delta r_{AC}(\theta_i)|$. Let $p$ be the pixel pitch of the projector on the focal plane, if $d_{BC} < d_{AC}$, a sufficient condition for the projections of points A, B and C fall into different pixels is $p < d_{BC} \cdot \sin(\Delta\theta)$. If not met, the trajectories of B and C will intersect at both $\theta_{i+1}$ and $\theta_n$, and the latter cannot be used to deliver $\Delta D_C$. In summary, angular resolution cannot be reduced indefinitely to increase the availability of projection angles. The inequality $\sin(\Delta\theta) > p/d$ is a guideline for choosing angular resolution, in which $d$ is the minimum distance between two points that are differentiable from all projecting angles, and can be used as a measure of the achievable Buildup resolution with ideal telecentricity.

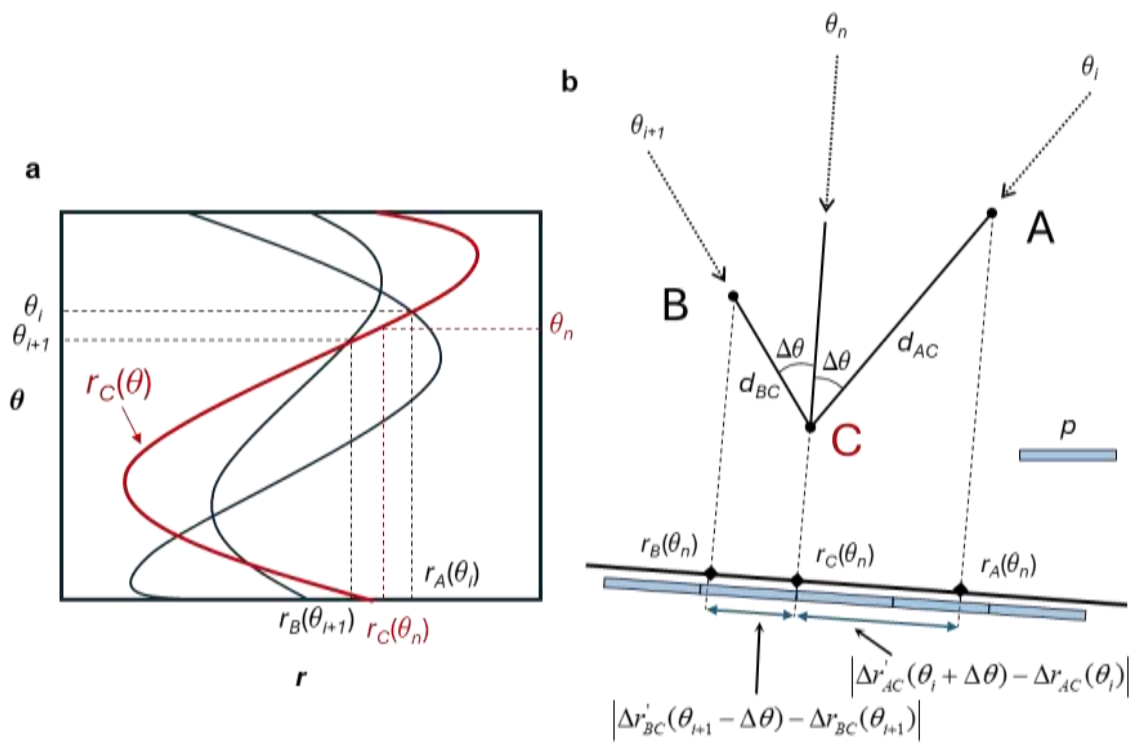

**Fig. 4**: **The advantage of dose subtraction can be compromised by limited availability of projection angles.** The total number of meaningful projection angles is limited by the resolution of the used projector. **a** Sinogram showing three trajectories of three points (A, B and C) in the curing volume postulated to define an effective angular resolution $\Delta\theta$. **b** An angular resolution is only meaningful if any displacement of a point's projection on the light source (the projector) caused by a stepwise rotation can be differentiated by the projector, i.e., falls into different bins defining pixels of the projector used.

**Periodicity**     To create strict photochemical negativity, i.e., to drive the motion of a system on the phase diagram without changing its distance to the origin (Fig.3a), the required irradiances depend on the instantaneous concentrations of A and B. This dependence entails that the gray value of a pixel in the sinogram for creating the same magnitude of negativity changes with time, material response to light, and varies from rotation to rotation. Practically, it is desirable to generate projections for a full rotation and repeat these projections periodically if multiple rotations are needed, and to account for variations in resin formulation by adjusting the output power of projectors and/or exposure time. If so, the doses a point receives, both that of the positive intensity that drags it closer to polymerization, and that of the negative intensity that pushes it away from polymerization, are fixed per full rotation. For an out-of-part point, the vertical displacement caused by UV light compensates for the traction towards polymerization caused by visible light, the self-decay of species B, and the consumption of species A caused by UV light, if any (Fig.5a). For an in-part point, the incident UV is minimized so that any vertical deviation from the x-axis can be off-set by visible light irradiation and self-decay (Fig.5b). The trade-off of this full-rotation based approach is that the magnitude of negativity scales with the output power of the UV light, and is qualitative rather than quantitative. For example, for an out-of-part point, its distance to origin may decrease over rotations, whereas in the case of strict negativity this distance should remain constant. Also, for an in-part point, a portion of the visible light is spent on counteracting the undesired UV light irradiation, which has implications for post-processing.

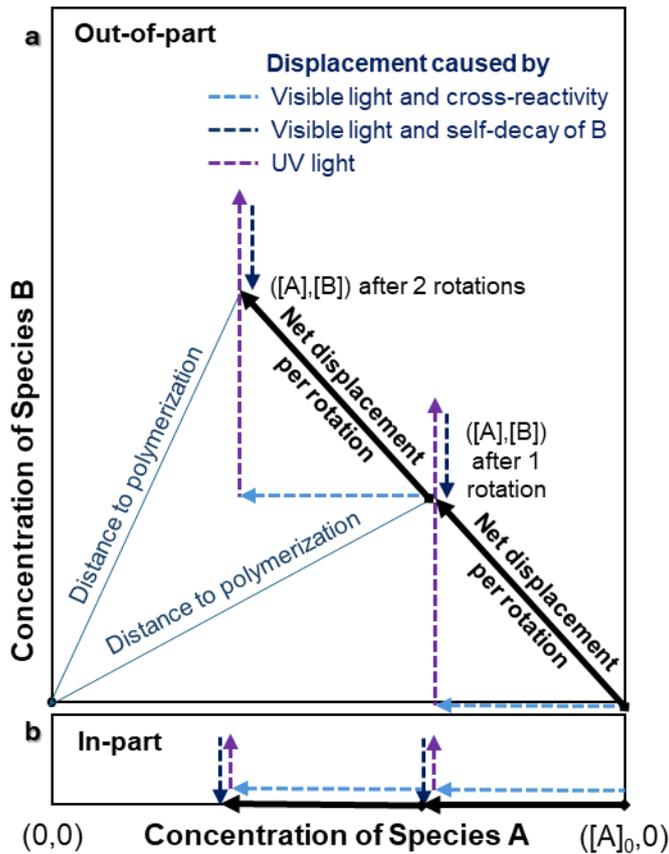

**Fig. 5: Binary photoinhibition (BPI) can be experimentally implemented based on full rotations of the curing volume.** Using the periodicity of rotations as the time unit has multiple advantages in projection optimization and experimental trial-and-error. **a** An exemplary path for an out-of-part point being steered on a full-rotation basis on the phase diagram. The distance to the origin may change within each rotation. The overarching goal is to use UV light to prevent the state of a system defined by ([A],[B]) from reaching the origin after a given number of rotations. **b** A typical path for an in-part point. A key requirement is that the vertical displacement caused by undesired UV dose can be compensated for by the visible light and the self-recombination of Species B.

Fig.6 shows how full rotation-based projections for a dual-color (DC) setup can be generated. The intensities of all pixels (either 0 or 1) are forward projected to create a green sinogram (Sinogram_G) with prescribed angular and spatial resolutions for visible light. The pixels are treated as discrete dots (with zero area), and the projections of two neighboring pixels on the sinogram are two individual points of zero width (instead of a line section). If we use "bins" to refer to projection pixels (i.e., discretization of a sinogram's $r$-axis), the points may fall into two nonadjacent bins if the spatial resolution of the sinogram is higher than that of the Design. If multiple points fall into the same bin, the width of which is defined by $p$, their intensities are summed. The width of the sinogram equals the Design's diagonal, and is determined by $p$ times the number of bins. Sinogram_G is then back-projected into a domain of the same resolution as the Design, producing a Buildup. The attenuation of visible light is accounted for using Beer-Lambert's law. The Buildup, after being rescaled to [0 1], is compared with the Design to produce a spatial distribution of dose oversupply (a matrix of the same size as the Design). This oversupply is forward-projected to create a sinogram (Sinogram_R) for UV light. Sinogram_R is then back-projected, with the attenuation of UV light being accounted for. The outcome is subtracted from the Buildup, after which the latter is rescaled to [0 1] and compared with the Design to yield a spatial distribution of dose undersupply. The undersupply is also of the same size as Design. If the root mean square of all the elements of this matrix is less than a set tolerance ($10^{-6}$), or the number of iteration exceeds a heuristic (23), the loop exists, and Sinogram_G and Sinogram_R are combined to produce the final sinogram. Otherwise, the undersupply pattern is forward projected to update Sinogram_G, entering the next iteration. A yellow pixel in the final sinogram indicates that both UV and visible light are enacted. Although this method does not guarantee exact reconstruction of the Design, it does not require sequential processing of pixels, nor does it rely on the availability of free projection angles. It is a generalization of our previous method for creating projections for two independent single-color setups (SC2) working in parallel.[19] The main difference being that the DC method remediates dose oversupply and undersupply separately by revising the sinograms of UV and visible light, respectively, whereas the SC2 method minimizes any mismatch between Design and Buildup by adjusting only one sinogram. The latter was made possible by setting a one-sided tolerance (e.g., 75% cut-off at maximum dose) for unintended dose buildup (i.e., over-exposure of out-of-part points in a curing volume), and the achievable contrast between in-part and out-of-part points can be affected by this tolerance. The MATLAB realization of DC and SC2 algorithms are supplemented as Script.S2 and S3.

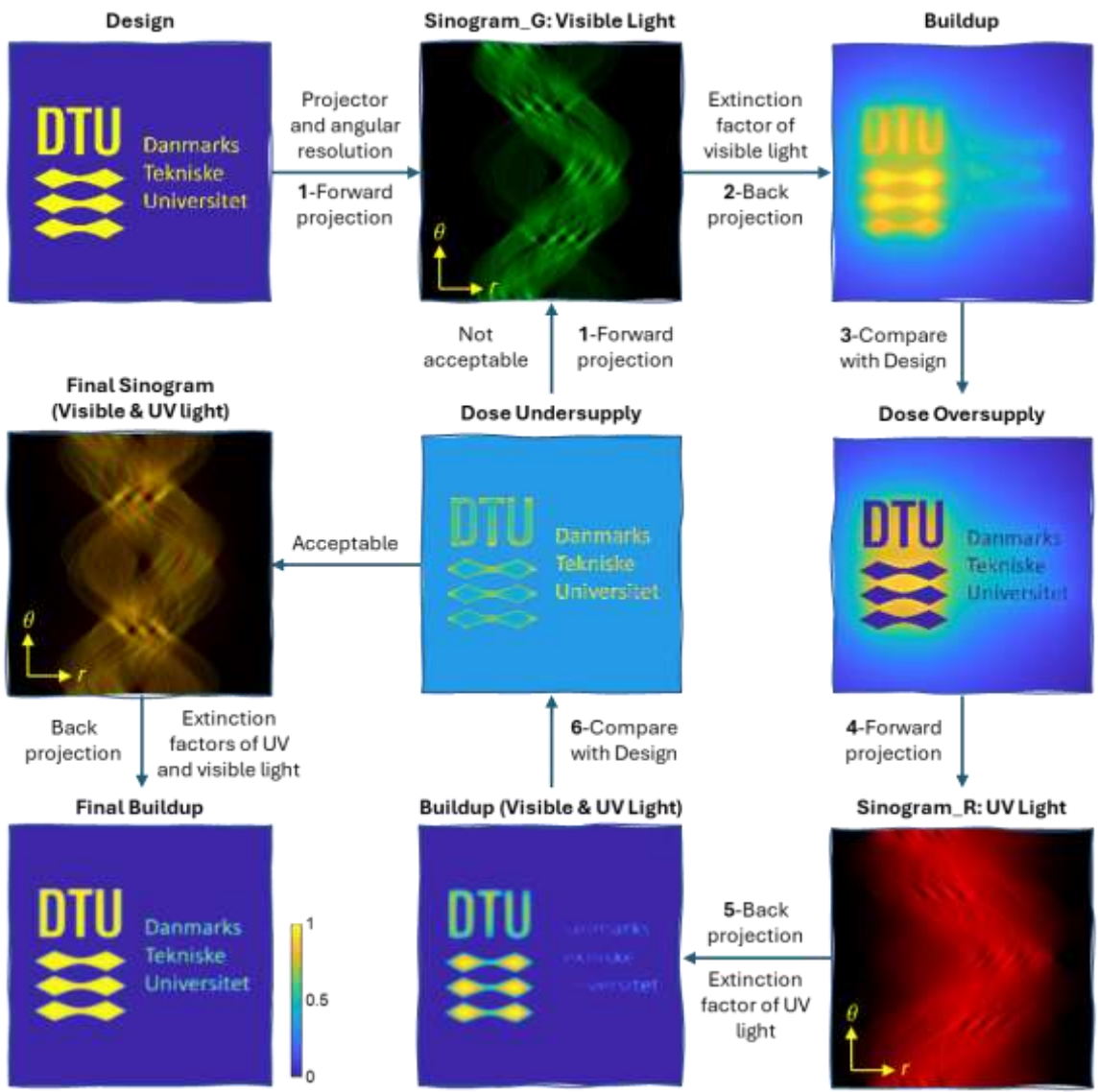

**Fig. 6: The sinograms for UV and visible light can be generated based on full rotations.** An iterative algorithm was developed for using visible light to supplement undersupplied dose and UV light to subtract oversupplied dose from an existing buildup. In the sinograms (Sinogram_G, Sinogram_R and Final Sinogram), green indicates visible light irradiation and red indicates UV light irradiation. Yellow indicates that both colors of light are projected

from the same pixel. In the Design and the Buildups, the color indicates relative dose within the lateral plane at different stages of the iteration. The Final Buildup was obtained by direct back-projection of both sinograms, and rescaled to [0 1]. See Script.S2 for MATLAB code.

**Printability** Given that the dose received by any point in the curing volume is constant per rotation, the achievable dose contrast between any pair of in-part and out-of-part points within a single rotation becomes critical for printability. For the polymer precursors used in this study, there are two important dose thresholds, $T_{opt}$ and $T_{mec}$ (Fig.7). The threshold $T_{opt}$ marks the initiation of photopolymerization and is determined by the initial concentration of oxygen. Reaching this threshold leads to a change in the optical characteristics of photoresin, which impairs the validity of assumptions built into sinogram generation. The threshold $T_{mec}$ is a heuristic and defines a desirable extent of conversion with which a workpiece stiffens sufficiently to sustain post-cure handling. Let $N_R$ be the total number of full rotations required to print an object, and $D_{in}$ and $D_{out}$ be the doses received by an in-part and an out-of-part point, respectively. For optimal printability, any pair of in-part and out-of-part points needs to meet three conditions: (a) $(N_R - 1)D_{in} < T_{opt}$, (b) $N_R D_{in} > T_{mec}$, and (c) $N_R D_{out} < T_{opt}$. Condition (a) prevents change in optical property before the final rotation and therefore minimizes detrimental effects such as self-lensing, (b) ensures that a workpiece is sufficiently stiffened for subsequential handling, and (c) prevents over-exposure. Not only is the difference between the doses $\Delta D$ important but also the absolute value of $D_{out}$, which is the main cause of over-exposure and cannot be shifted to zero by rescaling. Fig.7a shows a desirable scenario for a pair of points if the printing takes three full rotations. When the relative contrast is low ($D_{out}/D_{in} \to 1$), over-exposure occurs ($N_R D_{out} > T_{opt}$, Fig.7b). Downscaling the greyscale intensity or lowering the output power can alleviate over-exposure. However, as the relative dose contrast (the difference between irradiation doses received by two points in print-space, normalized to the maximum dose received by a point on the same lateral plane) per rotation for any point-pair is fixed, their absolute doses go down proportionally. As a result, such means of remediation can lead to under-exposure of in-part points (Fig.7c), compromising the structural integrity of a workpiece during post-processing. An alternative approach to improve printability without changing the sinogram is to prolong the induction by increasing the concentration of pre-dissolved inhibitors (e.g., by adding TEMPO). This is equivalent to combining conditions (b) and (c) to yield $N_R \Delta D > \Delta T$. Although $\Delta D$ and $\Delta T$ are fixed, the overall contrast is cumulative over time and improved printability can be achievable by increasing the number of full rotations (Fig.7e). In contrast, $D_{out}$ can be negative with BPI (Fig.7d), and condition (c) can be ignored. In other words, the key improvement in printability is that BPI enables building up very high dose contrast within a single rotation, including being able to achieve effectively $D_{out}/D_{in} \leq 0$. This advantage is further illustrated in Fig.8.

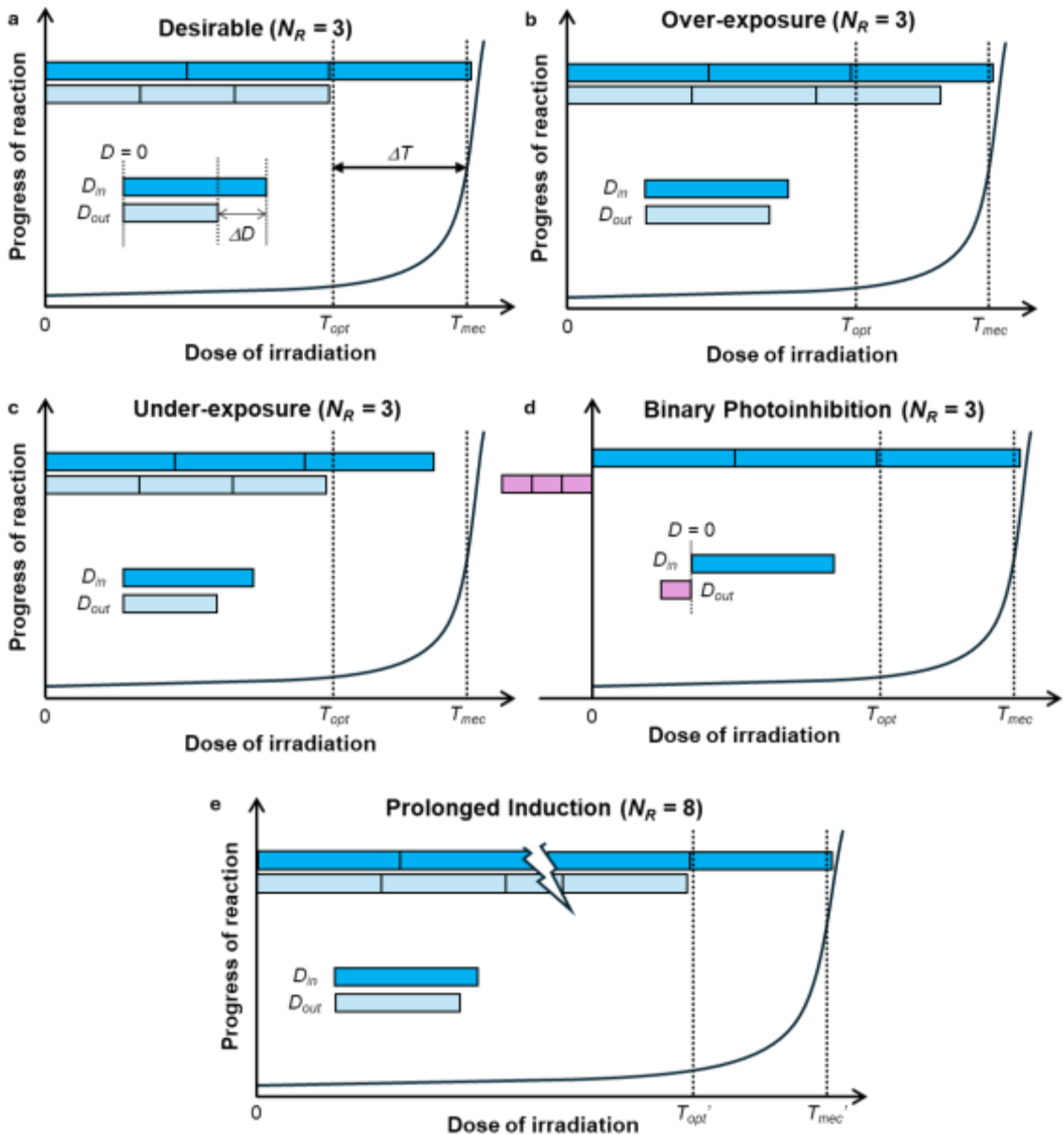

**Fig. 7: The printability of photoresins in tomographic volumetric 3D-printing reflects an interplay between dose delivery determined by sinogram and dose thresholds determined by choice of polymer precursors.** A high dose contrast ($\Delta D$), achieved via sinogram optimization and/or binary photoinhibition (BPI), will directly improve printability. **a** A desirable combination of sinogram and material response for an in-part and out-of-part point-pair. $D_{out}$ is sufficiently low to prevent polymerization of the out-of-part point after 3 full rotations, while $D_{in}$ is high enough to stiffen the in-part point for post-print handling. **b** Over-exposure happens when the out-of-part point polymerizes. **c** Under-exposure occurs when the extent of conversion at the in-part point was not sufficiently high. **d** An ideal implementation of BPI can maximize dose contrast by oppressing dose building in out-of-part

points. **e** Static inhibition improves printability by prolonging the induction period in which dose contrast per full rotation can be accumulated.

In a lateral plane, dose contrast is most difficult to establish for adjacent points close to the edge of a binary feature. Two generic requirements for $D_{in}$ are magnitude and uniformity. Combining printability conditions (a) and (b) gives $D_{in} > \Delta T$, i.e., the magnitude of in-part dose needs to be sufficiently large to "leap" from the left side of $T_{opt}$ to the right side of $T_{mec}$ (Fig.7a). This is in concordance with our observation that it is advantageous to use projectors with higher output power, although the benefit of the latter has more often been attributed to the minimization of detrimental effects caused by diffusion. Uniformity in $D_{in}$ for all in-part points is also desirable because it either prevents optical changes before the finalizing rotation, or minimizes the requirement for output power. Given a sinogram, let $D_{in,\min}$ and $D_{in,\max}$ be the lowest and highest dose received per rotation by in-part points, condition (a) requires that $(N_R - 1)D_{in,\max} < T_{opt}$ and (b) requires that $N_R D_{in,\min} > T_{mec}$. The magnitude of $D_{in,\min}$ scales directly with output power of visible light source, and is subject to $D_{in,\min} > (N_R - 1)(D_{in,\max} - D_{in,\min}) + \Delta T$. Because $\Delta T$ is determined by the polymer precursor, a low uniformity (i.e., large $|D_{in,\max} - D_{in,\min}|$) would indicate a greater power need for delivering $D_{in,\min}$, or the violation of (a). For optimal contrast, all $D_{out}$'s need to be minimized. These three requirements are difficult to meet simultaneously. Fig. 8a shows that the single-color Buildup attained from the algorithm outlined in Fig. 6 does not produce uniformity for in-part points, nor does it oppress the doses of out-of-part points at the edges of binary features. This low contrast is more pronounced when a Buildup is binarized using a threshold set by a given percentage of $D_{in,\max}$. For example, the Zoom-in panel in Fig. 8a shows that the counter of letter "a" is susceptible to over-exposure. When the Buildups of UV and visible light are superposed, however, good uniformity is achieved for both in-part and out-of-part doses, and an ideal overall contrast is achieved. The uniformity indicates that 0 and 1 are both achievable grey values with direct back-projection, which has the implication that BPI can be generalized to enable greyscale printing if optical changes accompanying $T_{opt}$ is negligible or can be accounted for in sinogram calculation. However, the high contrast is achieved at the expense of noticeable under-exposure at the edges of positive features. For example, zoom-in of the DC buildup shows that the strokes of letters are thinner than designed when binarized at 0.75 $D_{in,\max}$. Also, compared to its single-color counterpart, greater power of visible illumination would be required to achieve the same criterion set by $D_{in} > \Delta T$ in a dual-color setup, because much of the visible doses can be nullified by UV light. For comparison, in Fig. 8b we show a buildup using a previously reported algorithm,[19] with which it was possible to realize uniformity in in-part doses. A limited contrast can be established without the caveat of having reduced $D_{in}$ due to dose nullification. The strength of a dual-color system in maximizing relative dose contrast per rotation is evident in Fig.8c, in which buildups from mathematical reconstructions are binarized at 0.5 $D_{in,\max}$ for direct back projection (SC), binary photoinhibition (DC) and optimized single color projection (SC2). We note that Fig. 8 shows only mathematical reconstructions in an ideal telecentric setup, and is meant to be an indication of as-patterned quality on the lateral plane. The final quality of a workpiece can be impacted by the selection of $N_R$, the formulation of photoresin ($\Delta T$), the output power of light sources, and especially post-print handling.

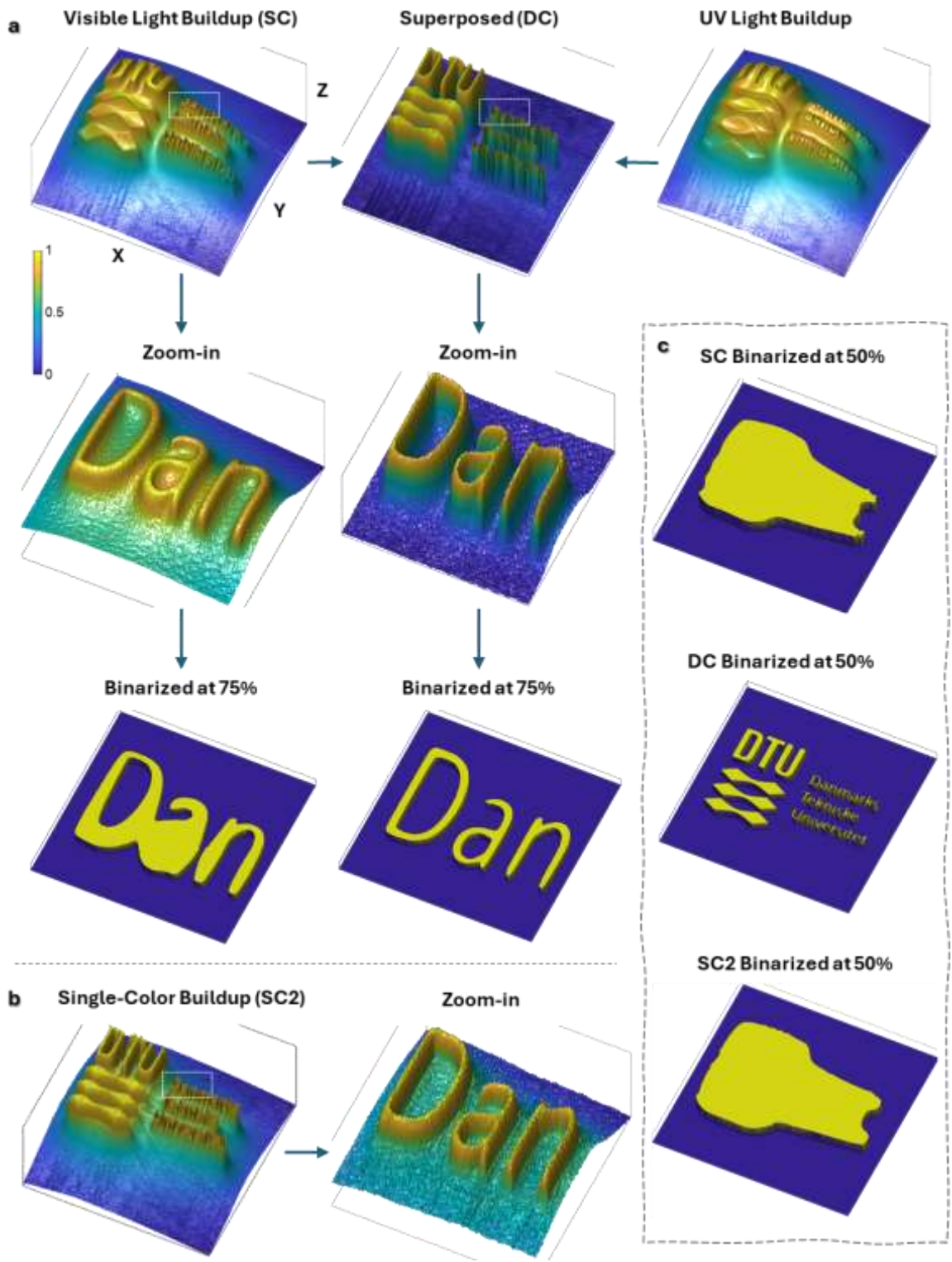

**Fig. 8: Superposition of dose buildups with binary photoinhibition (BPI) greatly improves achievable dose contrast within a lateral plane. a** Effective dose superposition via direct back-projection using sinograms generated on a full-rotation basis. Neither visible light nor UV light creates a buildup with high contrast between in-part and out-of-part features. However, when the UV light dose is subtracted from the visible light dose, the

effective buildup shows high uniformity and contrast. The relative contrast can be visualized by setting a binarization threshold at 75% of the maximum effective dose. **b** Without BPI, contrast can be enhanced only to a limited extent by optimizing the sinogram. The zoom-in shows details of the achievable contrast in a single-color setup when the background tolerance is set to 75%. **c** A comparison of achievable contrast: without BPI nor sinogram optimization (SC), with BPI (DC) and without BPI but with sinogram optimization (SC2). The colors indicate the relative magnitude of dose, rescaled to be between 0 and 1. The z-axis indicates the dose (corresponding to the greyscale colormap) or differentiates between in-part and out-of-part points after binarization.

**Experimental verification**

As an exemplary implementation of binary photoinhibition (BPI), we adopted a photoresin formulation using 2,2'-bis(2chlorophenyl)-4,4',5,5'-tetraphenyl-1,2'-biimidazole (*o*-Cl-HABI) as a source of secondary radical quencher.[33] Photolysis upon UV illumination cleaves an *o*-Cl-HABI molecule into two lophyl radicals, which inhibits free radicals-mediated polymerization because of the faster rate of radical-radical termination compared to radical-monomer interactions.[34, 35] *o*-Cl-HABI was chosen because of lophyl radical's relatively long lifespan and fair optical characteristics. The slow self-recombinatorial reaction of lophyl radicals leaves a lingering inhibitory effect[33] and is less energetically demanding for UV light to create negativity equivalence. In addition, we adjusted the concentration of *o*-Cl-HABI so that both visible and UV light can penetrate through the entire curing volume. It is, however, important to recognize that the oxygen-lophyl pair does not necessarily represent the full potential of ideal BPI (see Discussion). In addition, the concentration of *o*-Cl-HABI was deliberately kept low to optimize for light penetration, which limited the total amount of lophyl radicals available during printing. Nonetheless, with *o*-Cl-HABI, a dual-color system (DC-TVP) showed significant improvements over a single-color system (SC-TVP) in both as-patterned lateral fidelity and post-processed 3D prints.

We first used *in situ* shadowgraphy to detect as-patterned dose contrast within a lateral plane (Fig.9). Similar to projection optimizations,[6, 12, 26] the contrast enhancement with BPI is more pronounced in the as-patterned dose distribution, whereas the final quality of a workpiece may be affected by additional factors such as differences in material responses or post-print handling. Given the same starting material, shadowgraphy relies on local variations in refractive index to create gradients of light intensity on the detecting camera, and such variation stems from differed progress of photopolymerization. The latter is driven by the building of lateral dose contrast. Thus, *in situ* shadowgrams reveal the smallest distance between two points at which sufficient dose contrast can be established to produce a detectable variation in refractive index, and is therefore indicative of the lateral resolution for TVP. However, it is recognized that there is no direct relation between the gray value of pixels on the camera and the extent of conversion, and the method is qualitative only.

We first evaluated the effectiveness of BPI by patterning evenly spaced dots on a lateral plane. Square dot arrays (Fig.9a inset) of various density were patterned using both SC-TVP and DC-TVP. Two key differences were observed. First, for the single-color (SC) system, it was not possible to oppress polymerization between neighboring dots, and the discrete dot arrays appeared as interlaced grids of parallel lines (Fig.9a). For the dual-color (DC) system, there was a time window of up to tens of seconds in which the undesired polymerization between dots could be oppressed near the center of rotation (Fig.9b). Second, the smallest observable patterning distance between neighboring dots for a single-color system was ~162 μm, whereas for a dual-color system it was ~54 μm. When we further increased the dot density in DC-TVP, we observed periodically alternating gray values between neighboring pixels on the detector (each pixel corresponds to 27 μm on the imaged plane), suggesting formation of features beyond Nyquist criterion. We then tested more complex geometric designs, including a 4-group USAF resolution test target and an artistic design "Fall" that contains numerous curvy features. Both designs have an original resolution of 1500 by 1500 pixels, and the diagonal is aligned with the diameter of the cylindrical curing volume. With BPI, we managed to produce the three bars of all group elements from group -2 down to Element 4 of group 1 in the USAF target (Fig.9c). The formation of the six smallest bars (Elements 5 and 6) could be visually identified, but the resolution of imaging did not allow clear differentiation of the spacing between these bars (Mov.S2). We were not able to observe any comparable lateral patterning using a single-color setup because the time window in which features got blurred out due to over-exposure was brief. Also, larger and smaller elements did not form simultaneously, leading to co-existence of partial over-exposure and partial under-exposure within the same plane. For "Fall", the majority of curvy features were reproduced with good fidelity (Fig.9d, Mov.S3). The patterning of the DTU logo was less challenging due to its simplicity and larger spacing between features (Fig.9e), and continued illumination after the formation of said features did not lead to significant deterioration of the patterning quality (Mov.S4). We also tested a case of greyscale printing, here defined as delivering laterally patterned dose in continuous grey values. Unlike a single-color system, in which setting a tolerance for dose reception by out-of-part points is required in projection optimization, BPI allows effectively 100% dose contrast between in-part and out-of-part points. As we limit our discussion to binary printing in this work, we only briefly looked at the effect of BPI on the geometric fidelity of a workpiece in greyscale patterning (Fig.9f). For SC-TVP, if a greyscale design contains grey values lower than the set background tolerance, a convergence in iterative optimization cannot be achieved. For the SC2 method we adopted for SC-TVP, the tolerance is heuristically 75% of the in-part dose. Thus, the sinogram for the design in Fig.9f, which contains several dose grades below 0.75 $D_{in,\max}$, cannot be generated without ignoring these grades. BPI was not subject to such constraints, and the as-patterned geometric fidelity for a greyscale design with BPI far exceeded what would be achievable in a single-color setup. However, it is important to note that greyscale printing is typically used to control the extent of conversion in order to locally modulate the properties of a workpiece. Whether such a purpose can be served with BPI, or how severely the outcome will be impacted by post-print handling, remains unanswered using shadowgraphy. In summary, *in situ* shadowgraphy suggested that the introduction of BPI improves as-patterned dose contrast within the lateral plane, leading to an improved patterning resolution and good geometric fidelity of complex designs. However, whether these advantages can be fully maintained after post-processing cannot be warranted.

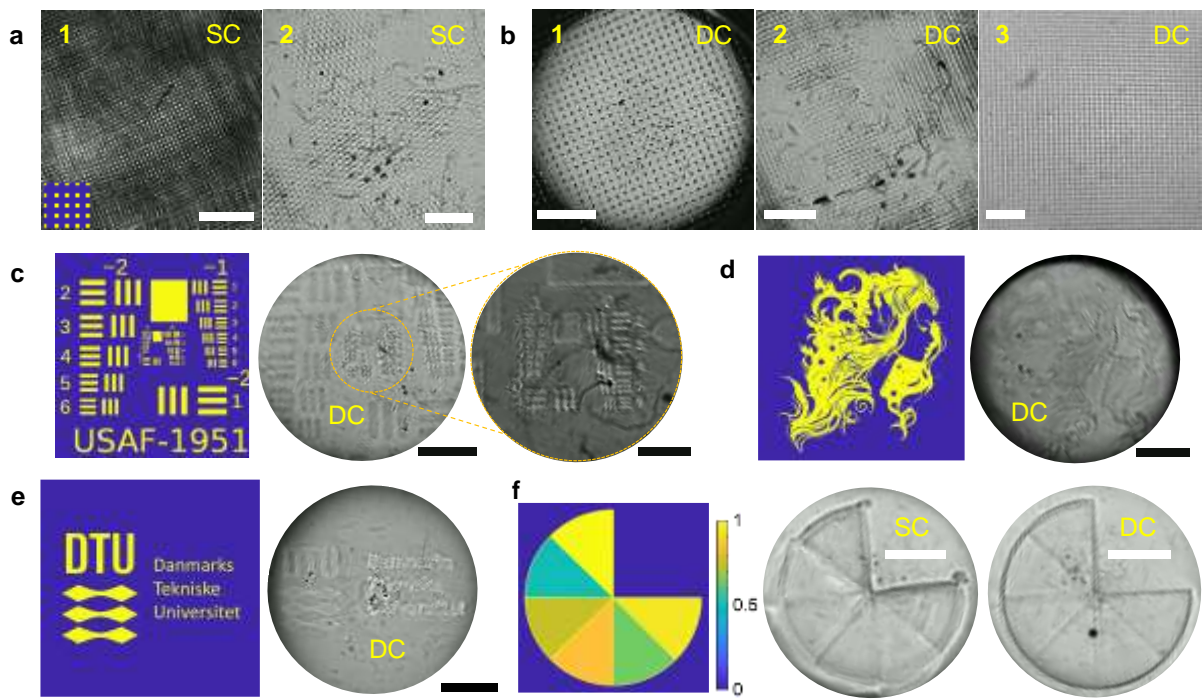

**Fig. 9: Binary photoinhibition (BPI) improves lateral patterning of geometric features.**
**a-b:** Lateral patterning of dot arrays (design pattern shown in **1a** inset) in a Ø30 mm curing volume using a single-color (SC, **a**) and dual-color (DC, **b**) setup, as captured using *in situ* shadowgraphy. Number of dots along the diameter: a1 – 800, a2 – 1300; b1 – 80, b2 – 800,

b3 – 1300. **c** Lateral patterning of an USAF target, zoom-in shows the smallest feature groups. See also Mov.S2 **d** Lateral patterning of a design with many curvy features. See Mov.S3. **e** Lateral patterning of the DTU logo. See also Mov.S4. **f** Geometric fidelity of greyscale patterning with and without BPI. The color indicates the relative grey value in the Design, scaled to [0 1]. Scale bars: a – 2 mm; b – 5 mm, 2 mm, 500 μm; c – 5 mm, 2 mm; d, e, f: 5 mm.

We then explored 3D-printing objects with various geometric characteristics with BPI. The Egyptian cat earring (Fig.10a) has a circular ring (~Ø2 mm) positioned vertically during printing. The main artefacts stemmed from vertical striation, which was observed using a scanning electron microscope (SEM) in many other workpieces, too. The spacing between two stripes are consistently ~20 μm in all imaged workpieces (Fig.S5), close to the nominal pixel pitch of the projector on the focal plane (15 μm). Fig.10b shows head of Michelangelo's David (National Gallery of Denmark). When printed using SC-TVP the main challenge was to preserve the shape of the nose and the curvy textures in the hair. Using DC-TVP, these features were less smeared out compared to its counterparts from SC-TVP. A miniature of the Eiffel tower was printed to evaluate the production of sharp edges and the synchronization of polymerization across the vertical direction. In a single-color setup, the termination of illumination is often determined by the vertical cross-section with the largest in-part area, which requires the longest exposure time to form. Polymerization may not occur simultaneously in all vertical positions if non-idealities in the actuation or light propagation are not fully accounted for, and the issue is more pronounced if a workpiece has vastly different cross-sectional areas vertically. The lack of such synchronicity requires specific termination of illumination at different vertical positions, which increases the operational challenges for TVP. When the tower model was printed using SC-TVP, the thinner tip always formed after the appearance of the square bases, leading to over-exposure near the bottom. In DC-TVP, the synchronicity issue persisted but to a much lesser extent. The order of vertical appearance was affected by the choice of maximum output power of UV light, and in the presented workpiece (Fig.10c) the entire tower was optimized to appear simultaneously in a sideview of *in situ* shadowgram. As a result, the print was terminated slightly prematurely (immediately after the tip was formed), leading to a mild collapse of a quarter of the square base (on the left in Fig.10c) during post-processing. The lateral sharp features were not difficult to produce within a few trial-and-errors using DC-TVP, as was also apparent in the printing of lattice blocks (Figs.10f-g). The internal compartments of the lattices can be washed out due to effective oppression by UV light, and the blocks can be seen through. Also presented in Movies S5-7 are full rotations of three workpieces with different compartment densities (all blocks have a lateral diagonal of ~30 mm). The workpieces are laid down (rotated 90 degrees) from its printing orientation so two of the four surfaces shown are lateral and two are vertical. To distinguish, all the lateral surfaces are square. We also tested designs with very fine surface features, such as a full body sculpture of *Frans af Assisi* (National Gallery of Denmark, Fig.10d) and a bust figurine of Albert Einstein (Fig.10e, Mov.S8). A comparison with *in situ* shadowgrams suggested that not all features produced during printing could be preserved through post-cure handling, even with meticulous trial-and-errors. It is important to recognize that three key elements are essential to the final quality of a workpiece: termination of illumination, choice of the ratio between UV and visible light's output power, and sample handling during post-processing. These elements are intrinsically correlated and were optimized for each 3D design individually. In this work, a holistic decision on how to tweak these elements coherently for an optimal outcome was not entirely objective and relied on the experiences of the operators. Therefore, even though we believe that BPI introduces clear advantages to TVP, it was impractical to fully quantify such advantages given the uncertainties introduced by the subjectivities. Below we instead give a quantitative analysis of the impact of UV irradiation on lateral feature formation.

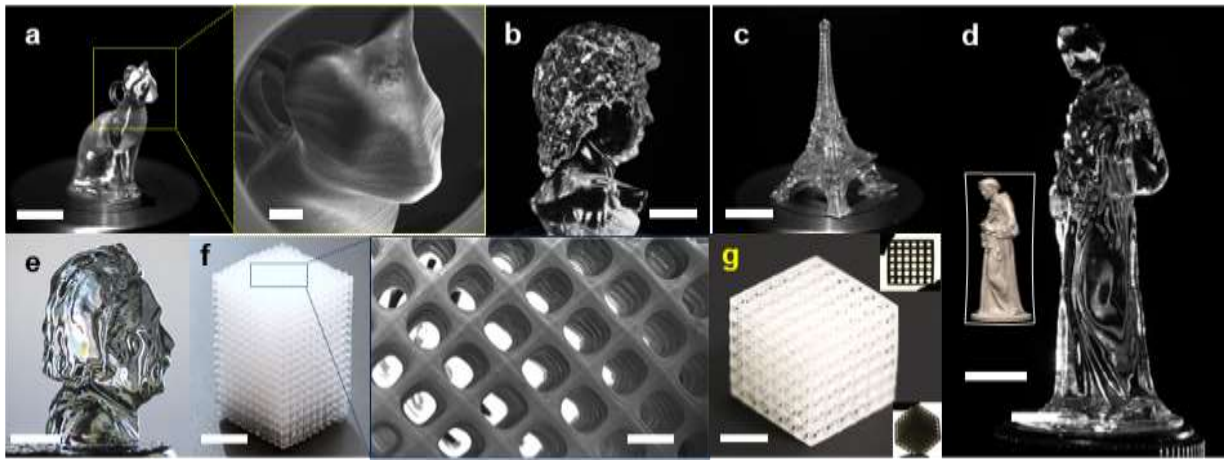

**Fig. 10: Physically printed objects using a dual-color tomographic volumetric 3D printer. a** Egyptian cat earring. **b** David's head by Michelangelo (National Gallery of Denmark, Copenhagen, Denmark). **c** An Eiffel Tower model. **d** Frans af Assisi (National Gallery of Denmark, Copenhagen, Denmark). **e** Albert Einstein Bust. **f** a 10×10×16 lattice block with patent interior. **g** a 6×6×6 lattice block (insets show the hollowness of the same workpiece). Scale bars: a – 10 mm, 1 mm; b, c, d, g – 10 mm; e – 2 mm; f – 3 mm, 500 μm.

Fig.11 shows a compilation feature sizes for 5 lattice blocks printed with BPI. Workpieces W1 and W2 contain 13×13×12 cubic compartments and W3-W5 contain 5×5×6 cubic compartments (Fig.11a). The shown features were located at the center of each imaged plane, and were used for feature size measurements (Fig.11b). The same output power of visible light (~65 mW/cm$^2$) and similar overall exposure (3 to 4 full rotations in 2 minutes, manually terminated based on visual inspection on *in situ* shadowgraphy) were used, while the output power of UV light were changed by setting the operating current of the LED bulb. The number of rotations was chosen to avoid visually identifiable over-exposure, which happened quickly after 4 full rotations for all workpieces and would clog many of the compartments. All workpieces went through the same post-print handling and were coated with nanometer-scale carbon before being imaged by a scanning electron microscope (SEM). Visual inspection revealed that the uniformity of the square features on the vertical plane is superior to that of the lateral plane, with the latter typically showing radial dependence, i.e., features closer to the edge of curing volume were more likely to over-expose. On the lateral plane, the square features on W3, W4 and W5 were more uniform than those on W1 and W2, which we attributed to the simpler geometric design and the greater spacing between positive features. The positive features were measured by counting the distance between two pixels at the narrowest part of a strut, and the negative features were the longest distance normal to and between two neighboring parallel struts. The uncertainties were calculated from multiple measurements from the same image (raw data compilation: Supplementary Data 01). On the lateral plane, features along both X and Y directions are measured. On the vertical plane, only features aligned vertically during printing were measured. The orientation was identified macroscopically by the number of compartments along each dimension and microscopically by identifying striation such as the ones shown in Fig.S5. In general, 3 full rotations led to under-exposure of positive features. However, because it is typically more challenging to oppress over-exposure for negative features than to prevent under-exposure, especially on the lateral plane, we consider these results relevant and expect that the accuracy of fabrication can be improved by fine-tuning the combination of exposure time, number of rotations and the output power of both wavelengths. It is noteworthy that the measurement of the vertical features for W2 was affected by the detrimental effect of surface charging on the contrast of the image. Overall, the uncertainties of the vertical features were less than those observed on the lateral plane, and the power of UV light showed a more pronounced impact on the formation of lateral, negative features. Both observations conform to the different contrast-building mechanisms and corroborate the effectiveness of BPI in controlling lateral feature formation.

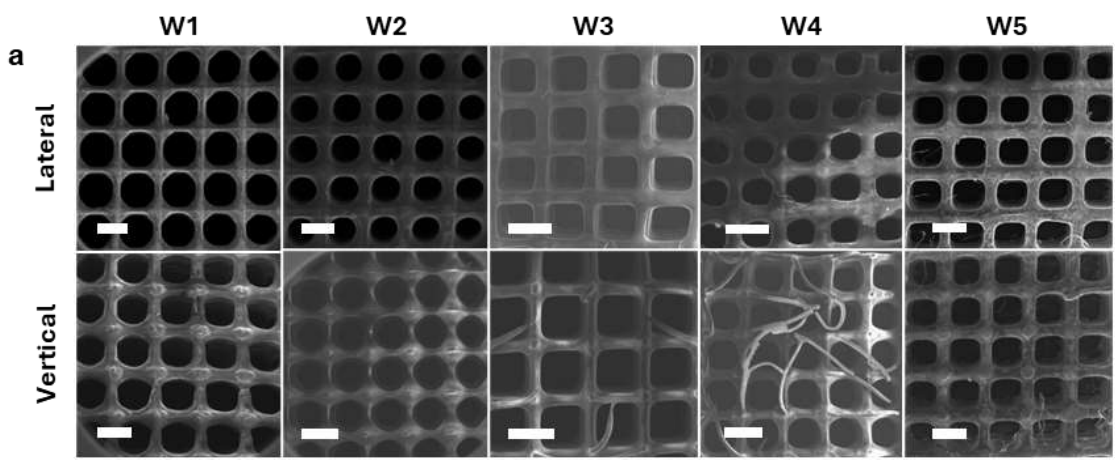

| Workpiece | $P_{UV,max}$ (mW/cm$^2$)* | Plane | Positive Features Design | Positive Features Measured | Negative Features Design | Negative Features Measured | Measured Ratio** |
|---|---|---|---|---|---|---|---|
| W1 | 45.6 | Lateral | 375 μm | 252.0 μm ± 39.3 μm | 750 μm | 1162.2 μm ± 49.5 μm | 4.61 |
|  |  | Vertical | 375 μm | 300.1 μm ± 24.6 μm | 750 μm | 778.1 μm ± 30.8 μm | 2.59 |
| W2 | 22.8 | Lateral | 425 μm | 399.7 μm ± 47.5 μm | 850 μm | 806.5 μm ± 39.8 μm | 2.02 |
|  |  | Vertical | 425 μm | 377.6 μm ± 17.2 μm | 850 μm | 661.1 μm ± 27.4 μm | 1.75 |
| W3 | 45.6 | Lateral | 450 μm | 279.4 μm ± 36.2 μm | 900 μm | 871.4 μm ± 46.9 μm | 3.12 |
|  |  | Vertical | 450 μm | 233.1 μm ± 8.3 μm | 900 μm | 946.7 μm ± 42.3 μm | 4.06 |
| W4 | 22.8 | Lateral | 450 μm | 310.9 μm ± 27.0 μm | 900 μm | 684.1 μm ± 45.0 μm | 2.20 |
|  |  | Vertical | 450 μm | 237.2 μm ± 24.9 μm | 900 μm | 817.2 μm ± 30.7 μm | 3.45 |
| W5 | 34.2 | Lateral | 450 μm | 324.1 μm ± 37.4 μm | 900 μm | 921.0 μm ± 42.2 μm | 2.84 |
|  |  | Vertical | 450 μm | 326.0 μm ± 29.4 μm | 900 μm | 904.6 μm ± 50.4 μm | 2.77 |

*Maximum UV power on the focal plane, given the same output power of visible light (~65 mW/cm$^2$).
**Negative-to-positive feature ratio, the design value for all workpieces is 2

**Fig. 11: 3D-printed lattice blocks show the impact of UV light on the formation of lateral features.** The post-print handling of the five workpieces were kept constant to the best extent possible. Same output power of visible light and similar overall exposure time were used. **a** Lattice features on the lateral and vertical planes were analyzed separately using secondary electron imaging. Although only the surface of each block was imaged, visual inspection confirmed that all the interior compartments were hollow and of similar dimensions. **b** Both positive and negative features were analyzed. Statistics were obtained from measuring multiple features from the same image. Scale bars: 1 mm.

The difficulty in identifying an optimal number of rotations points to several practical challenges in implementing binary photoinhibition. First, cross-reactivity, in this context defined as the potential mismatch between absorptivity and efficiency of intended photochemical reactivity[36], and the ability of UV light to initiate photopolymerization. The emission spectrum of the UV light source extends beyond 405 nm (Fig.12a), and in the absence of *o*-Cl-HABI triggers double-bond conversion given sufficient exposure (Figs S6-7). This cross-reactivity issue is less pronounced when *o*-Cl-HABI is abundant, as the inhibitory effect of lophyl radicals significantly outweighs the photoinitiation. However, at a low concentration, which was required in this study for optimal resin transparency, *o*-Cl-HABI may deplete during a full rotation, which voids the validity of pre-calculated dual-color projections: UV light no longer nullifies visible light, but instead promotes the polymerization of both in-part and out-of-part points, leading to rapid over-exposure (Fig.12b). Meanwhile, before depletion, the effective dose buildup in in-part points is reduced by undesirable UV illumination, and therefore the printability criterion $D_{in} > \Delta T$ is more difficult to meet in a DC-TVP (Fig.12c). As a consequence, it was more difficult to sustain the structural integrity of prints with BPI during post-print handling, and structures formed with low extent of conversion could be washed away in post-processing. This reduction in $D_{in}$ contributed to the apparent under-exposure of the workpieces presented in Fig.11. The issue can be alleviated if the onset of optical changes at $T_{opt}$ is negligible (thus avoiding the need for $D_{in} > \Delta T$), and the number of rotations can be increased given sufficient *o*-Cl-HABI.

Overall, practical implementation of BPI can be susceptible to detrimental effects of cross-reactivity, limited availability of the secondary inhibitory species, and weak structural integrity. These challenges contributed to the difficulty of preserving fine features observable in *in situ* shadowgraphy through post-processing.

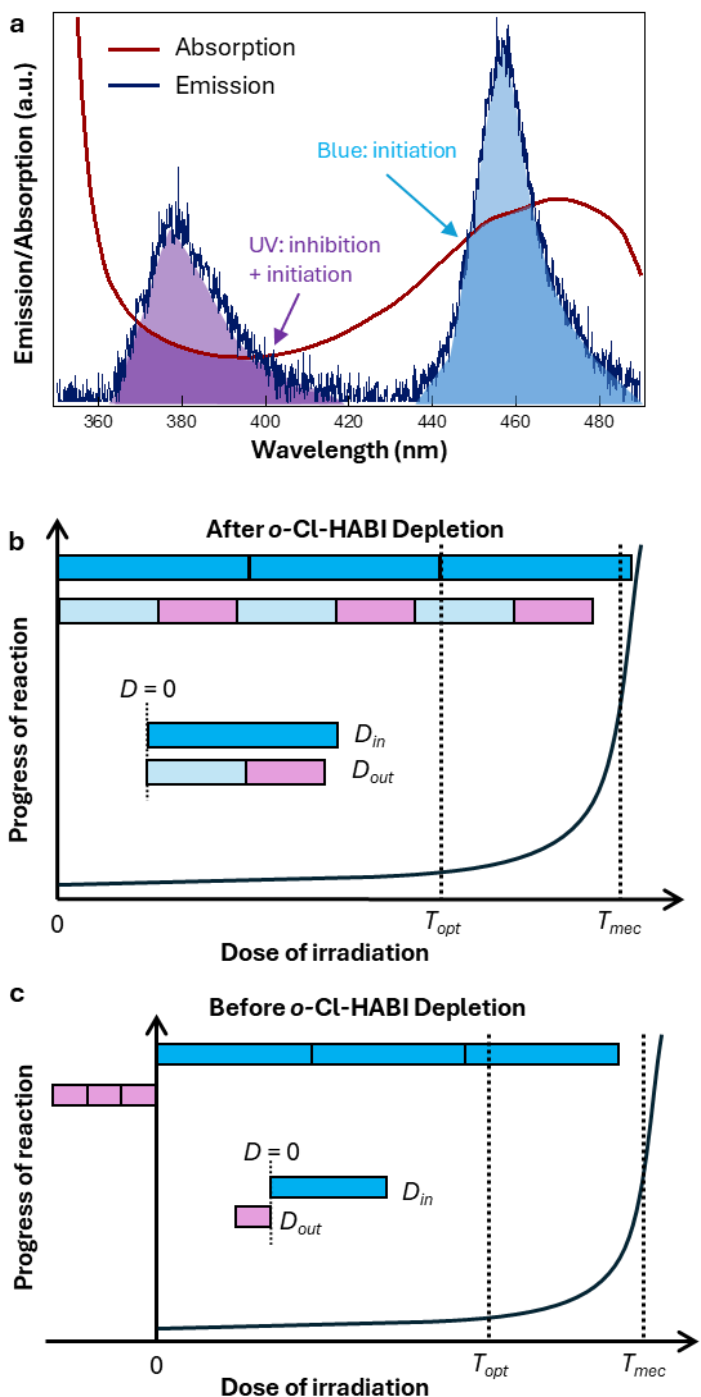

**Fig. 12: Limitations of the oxygen-lophyl pair include wavelength specificity, lophyl radical availability and structural integrity. a** Normalized emission spectrum of the projector and absorption spectrum of photoresin. **b** After the depletion of $o$-Cl-HABI, the free radicals generated by UV will exacerbate over-exposure of out-of-part points. **c** The extent of conversion of in-part points is generally lower due to the inhibitory effect of lophyl radicals, leading to challenges in post-print handling.

**Discussion**

When bisphenol A glycerolate dimethacrylate (bisGMA) and triethylene glycol dimethacrylate (TEGDMA) are mixed at a 2:1 weight ratio, the viscosity of mixture was measured to be 18.8 Pa·s. This viscosity was not sufficient to completely prevent sedimentation of workpiece during a long printing process. As remediation, when investigating lateral feature formation, we printed large but thin slices embedded with features shown in Fig.9. The slices laterally span the entire curing vial but are of less than 1 mm thick. The formation of features were recorded in real time (Fig. 9, Mov. S2-4), and no noticeable effect of sedimentation was observed throughout the process. For 3D objects, the sedimentation could be more pronounced. We used flat-bottom vials so that we could build objects directly sitting on the bottom of a vial (Mov. S8). For hollow structures, the exposure time could be further shortened because the net volume of polymerization was small. For example, the lattice blocks in shown in Fig. 11 were printed in ~2 min at the bottom of vials to minimize the detrimental effects of sedimentation.

The viscosity was used in the Wilke-Chang equation[37] to estimate the diffusivity of inhibitors

$$D = 7.4 \times 10^{-6} \sqrt{\psi \overline{M}} \frac{T}{\mu} \left( \frac{\rho}{M} \right)^{0.6} \quad (11)$$

Here $D$ is the diffusivity (mm$^2$/s), $\psi$ is an association parameter for the solvent (assumed to be 1), $\overline{M}$ is the average molecular weight of the photoresin (438 g/mol), $T$ is temperature (293.15 K), $\mu$ is the viscosity of the resin (18,800 cP), $\rho$ is density of the inhibitor at boiling point as liquid (1.14 g/mL for oxygen and 0.88 g/mL, the value for benzene, was used for lophyl radical), and $M$ is the molecular weight of the inhibitor (32 g/mol for oxygen, 330 g/mol assumed for lophyl radical). Eq. 11 estimates that $D_{oxygen} = 3.27 \times 10^{-7}$ mm$^2$/s and $D_{lophyl} = 6.89 \times 10^{-8}$ mm²/s. Even though Eq. 11 serves only as a rough estimate, it is expected that the diffusivity of lophyl radicals is lower than that of oxygen because of the difference in molecule weight. In TVP, the spatiotemporal distribution of inhibitors defines the shape of a workpiece[12, 19]. The diffusion of inhibitors thus affects the formation of fine features. The evolution of inhibitor gradient reflects a concerted interplay between light propagation, chemical reactions, and diffusion of reactants. Mov. S9-10 visualize spatiotemporal distribution of oxygen, o-Cl-HABI and lophyl radicals in a reactive transport simulation. The coloration indicates the relative magnitude of a visualized quantity with blue being zero and red being one (the "jet" colormap in MATLAB). Mov.S9 shows a simulated dual-color printing process in which depletable o-Cl-HABI and effect of its evolving concentration on light extinction coefficients have been accounted for. Mov.S10 shows the evolution of the BPI for two points of interest (A – out-of-part, B – in-part). For comparison, Mov.S11 shows the same geometry printed using SC-TVP. Givn the complexity of such processes, it is hard to determine the smallest printable feature solely from diffusivities. However, the diffusion length ($L_D$) in two limiting cases is particularly relevant for BPI. We can use $L_D = \sqrt{6D\tau}$ to estimate the distance inhibitors diffuse within the characteristic time of a said process[12], in which $D$ is the diffusivity (mm²/s) and $\tau$ is the characteristic time (s). When oxygen is abundant, diffusion of oxygen dominates the transport of inhibitors, and the characteristic time corresponds to that of the full elimination of oxygen by the primary writing wavelength. This can be the typical time for a single-color printing process (e.g., $\tau_{SC}$ =100 s)[12], and $L_{D,oxygen} = \sqrt{6D_{oxygen}\tau_{SC}} = 1.40 \times 10^{-2}$ mm. In the other limiting case, oxygen is depleted, and the diffusion of lophyl radicals dominates fine feature formation. In this case, the characteristic time is the lifespan of lophyl radical, subject to the eradication by the primary writing light and the self-recombinatorial reaction to form o-Cl-HABI. It is reported that the polymerization rate of methacrylate recovers approximately 30 seconds after the cessation of UV irradiation[33], which was corroborated by *in situ* measurements of double bond conversion with dual-color illumination (Figs S6-7). Therefore, if we choose $\tau_{DC}$ to be 30 seconds, $L_{D,lophyl} = \sqrt{6D_{lophyl}\tau_{DC}} = 3.52\times10^{-3}$ mm, significantly less than $L_{D,oxygen}$. Overall, even though DC-TVP with BPI may require longer printing time than SC-TVP, we do not expect additional detrimental effects stemming from inhibitor diffusion because of the depletion of oxygen and the low diffusivity and limited lifespan of lophyl radicals.

The oxygen-lophyl pair differs from an ideal BPI system in that the total amount of available lophyl radicals is limited by the concentration of *o*-Cl-HABI. As discussed in the "Binary Photoinhibition (BPI)" section above, a prerequisite for pinning a system indefinitely at a stationary state other than the origin is an unlimited supply of Species B. An abundant supply does not guarantee more efficient BPI (Fig.S8), as the lifespan of Species B can become a bottleneck. In this study, we found that 0.4 wt% *o*-Cl-HABI sufficed to generate lophyl radicals that clearly demonstrated the advantages of BPI. Higher concentrations led to stronger light extinction (Fig.S9) that would deteriorate workpiece fidelity towards the rotation center. At the chosen concentration, UV irradiation at 70 mW/cm$^2$ for 15 seconds did not deplete *o*-Cl-HABI (Fig. S10). Also noteworthy is that the lophyl recombination reaction did not follow an apparent first order law and therefore did not have a constant half-life. However, its decay is sufficiently slow to be cumulative during printing. The generation of lophyl radical follows a zeroth order kinetic with respect to UV irradiance (Fig. S10). It remains unclear how the competition between oxygen and lophyl radicals for visible light scales with their relative concentration. However, the prolonged induction period due to presence of lophyl radicals varied according to the timing of introducing UV (Fig. S7). When 30 seconds of UV illumination was introduced before visible light, the induction period was prolonged from 30 s to 38 s. When the same UV illumination was introduced after 30 seconds of visible light illumination, the induction period was instead prolonged to 50 s. The latter represents a case with a relatively lower oxygen-to-lophyl ratio, and the extended induction indicates that the consumption of visible dose by lophyl radicals does correlate with its relative concentration to oxygen – in the second case, more lophyl radicals were consumed and thus less was lost to self-recombination. If co-illumination were employed to minimize the effect of self-recombination, and the inductive period would be further prolonged to 68 s.

In this study, post-processing had been a non-negligible source of uncertainty. Objects produced with BPI were in general softer due to lowered conversion, and required more skillful post-print handling. In a typical process, the entire curing volume (the test tube containing processed photoresin) was heated to reduce the viscosity of photoresin, after which the residual resin was poured out from the tube. The tube was then filled with isopropanol alcohol (IPA) and capped, shaken gently, then flashed with unstructured UV light of 405 nm for 10 seconds. This pre-hardening step could facilitate following transport steps but could also cause clogging of enclosed features (e.g., internal compartments in lattice blocks). We also observed compromised surface finish if a longer pre-hardening was used. A hot air blower was used to dry the IPA on the surface. A dried workpiece was put in a stream of nitrogen and flashed by unstructured UV for an additional 10 minutes. This procedure helped preserve many fine features produced by BPI, but each step involved certain subjectivity in sample handling. We recognize that a more standardized post-processing protocol would improve the reproducibility of high-quality prints.

In summary, this work presents a new method to improve TVP's print quality by designing photoresponses using multiple wavelengths. We observed that building dose contrast in a lateral plane is a key challenge in developing TVP, which has previously been tackled via projection optimization. We showed that dose subtraction and a sufficient number of projection angles theoretically enables exact reconstruction of a greyscale design on a lateral plane. Dose subtraction can be realized via binary photoinhibition (BPI). We employed oxygen - lophyl pair to demonstrate the control of lateral feature formation using DC-TVP. *In situ* shadowgraphy showed that BPI was effective in enhancing as-patterned lateral dose contrast, producing localized differentiation in refractive index within 54 μm. Although not all fine features created during dose patterning could be preserved through post-processing, BPI still showed qualitative improvements in the fidelity of 3D workpieces. We then showed quantitatively that UV irradiation had a direct impact on the formation of negative features on a lateral plane, while the effect was less pronounced on the vertical plane, conforming the difference between contrast-building mechanisms. Looking forward, we recognize the importance of developing a BPI with less pronounced cross-reactivity, longer lifespan and greater availability of the secondary inhibitor, and improved optical properties (lower extinction and minimum coloration upon stimulation). Finally, we argue that a post-processing protocol that maximizes workpiece reproducibility would be essential for moving TVP research closer to industrial applications.

**Materials and methods**

**Materials**    The photoresin was prepared by mixing triethylene glycol dimethacrylate (TEGDMA, CAS#109-16-0, Sigma–Aldrich) and bisphenol A glycerolate dimethacrylate (bisGMA, CAS#1565-94-2, Sigma–Aldrich) at a weight ratio of 1:2 for optimal viscosity. The viscosity of this formulation was measured to be 18.8 Pa·s using a Discovery Hybrid rheometer (TA Instruments, USA). The TEGDMA was shipped with 80-120 ppm 4-methoxyphenol as inhibitor, the effect of which on BPI has been ignored in this study. Camphorquinone (CQ, CAS# 10373-78-1, ≥96.5% purity, Sigma–Aldrich) and ethyl 4-dimethylaminobenzoate (EDAB, CAS# 10287-53-3, Sigma–Aldrich) were added as the photoinitiator (0.1 wt%) and co-initiator (0.25 wt%), respectively (Fig.S11). 2,2'-Bis(2chlorophenyl)-4,4',5,5'-tetraphenyl-1,2'-biimidazole (*o*-Cl-HABI, CAS# 7189-82-4, TCI Europe) was first dissolved in tetrahydrofuran (THF, CAS# 109-99-9, Fisher Scientific) at 28 wt% then added to the photoresins at 0.4 wt% (Fig.S12-13). The absorption spectrum of this photoresin formulation is shown in Figure 12a in the context of the emission spectrum of the projector. The refractive index of the photoresin was measured as a function of wavelength (Fig.S14). All the chemicals were obtained commercially and used as received without further purification.

**Dual-Color Tomographic Volumetric 3D Printer** Fig.S15 shows a schematic drawing of the 3D-printer used in this study. Component S1 is a dual wavelength 4K DLP projector (TVP07-15, Xiamen Zhisen, China) equipped with both blue and UV light sources (Fig.12a). It was used with a bi-telecentric lens set (L1, Xiamen Zhisen, China) to approximate parallel beam relay into the curing volume. The lens system was designed to compensate for the achromatic aberration of 385 nm and 460 nm. Irradiance was measured as a function of greyscale intensity at these wavelengths (Fig. S16). Two *in situ* shadowgraphy imaging systems were integrated in the printer. The axial imaging system was illuminated by an LED source (S2, $\lambda$ = 625 nm, M625L4-C4, Thorlabs) through a beam expander (BE, D series 5x, Wuhan Baixin, China) to ensure collimated light covers the entire curing volume. A telecentric lens (TL2, ESCM014-180X23, ES technology, China) was used to demagnify the curing volume to the imaging size of the detector (C1, CMOS camera, GS3-U3-51S5M-C, FLIR Grasshopper). The lateral imaging system used a collimated light source (S3) that could directly illuminate the curing volume in the lateral direction, followed by a 4f lens system composed of lens L1 (200 mm, LA1979-N-BK7, Thorlabs) and L2 with an adjustable working distance (25 mm, TECHSPEC59-871, Edmund) to relay the beam to camera C2 (same as C1). The maximal projection area in the focal plane was 32.4 by 57.6 $mm^2$, with a nominal pixel pitch of 15 μm. An in-house mounting setup was employed to mount flat-bottom cylindrical test tubes (Ø30 mm). Two 60 mm self-centering lens holders, CRD-XA (Oeabt, China), connected via four 6mm PCM-S series rods, were combined into a gripper. Lens holders were assembled on an X-Y manual linear translation stage, PT-XY125 (PDV Instrument Co., China), with a resolution of 10 μm. The rotation stage was calibrated using this setup to eliminate misalignments in the test tubes during rotation. All components were mounted on a motorized rotation stage, 8MR190-2 (Standa, China). A cuboid vat containing index-matching fluid was placed outside the test tube, with walls perpendicular to incident beams.

**Optical Characteristics of the Projecting System** A complementary metal oxide semiconductor camera (FLIR Grasshopper GS3-U3-5185M-C CMOS) was used to characterize the modulation transfer function (MTF) of the dual-color projector at various positions across the curing volume. The detector has a nominal pixel pitch of 3.45 μm and a native resolution of 2448×2048. A 2160×3840 slanted-edge image was projected and captured on 5 planes (15, 6, 0, -6, -15 mm from the focal plane). The recorded images were processed using ImageJ to obtain the MTF values with the SE_MTF_2xNyquist plugin (Fig.S17).

**Attenuated Total Reflectance - Fourier Transform Infrared spectroscopy (ATR-FTIR)**
ATR-FTIR (Spectrum 100, PerkinElmer, UK) was used to measure double bond conversion. A 1.5 mm ring was used as a sample holder, placed next to the ATR cell of the spectrophotometer. A pipette was used to control the sample volume and transfer the test subjects. The spectra were obtained from 4000 $cm^{-1}$ to 650 $cm^{-1}$ with one scan per 8.3 or 4.3 seconds. Real time conversion was calculated by monitoring the C=C alkene stretch peak at 1638 $cm^{-1}$ and that of the aromatic ring at 1610 $cm^{-1}$. During *in situ* measurements, a sample was illuminated using 470 nm blue LED (SOLIS-470C, Thorlabs) and/or 365 nm UV light (M365LP1-C1, Thorlabs). The output power of the LEDs was set via LED drivers, and irradiance was measured prior to each experiment using a standard photodiode power sensor (S120VC, Thorlabs) working in the 200 - 1100 nm wavelength range and recorded with an optical power console (PM400, Thorlabs).

**UV-Vis Spectroscopy** UV-Vis spectroscopy was performed with an Agilent Instrument Exchanges Service Model G1103A using a 10 mm path length quartz cuvette to measure the absorption character of the chemical reagents. The absorbance spectrum was collected from 200 nm to 900 nm. As a reference, the absorption characteristics of the solvents were also measured. The extinction coefficient of the reagents was calculated using the Beer-Lambert law. *In situ* UV-Vis measurement was used to evaluate the evolution of the lophyl radical concentration upon the photolysis of *o*-Cl-HABI. In such tests, UV and visible light sources were placed on the opposite sides of the cuvette. In order to remove pre-dissolved oxygen molecules, each sample was first illuminated by visible light for 15 seconds at 5 mW/cm$^2$, then by UV for 20 s. The *in situ* absorbance was recorded every 6.9 s. All measurements were carried out in a dark environment.

**Viscosity Measurements** The viscosity of photoresin was measured using a Discovery HR-2 Hybrid rheometer (TA Instruments, USA) equipped with 40 mm diameter parallel plates. A constant plate gap of 1000 μm was used. Measurements were conducted across a shear rate range of $10^{-3}$ to $10^3$ s$^{-1}$ at 25 °C on a temperature-controlled plate.

**Post Processing** In a typical process, the test tube containing the workpiece and liquid resin was heated to reduce the viscosity of photoresin, after which the residual resin was removed. Isopropanol alcohol (IPA) was then injected into the tube, which was then capped, gently shaken and UV (405 nm) flashed for 10 seconds before the workpiece is picked out. A hot air blower was used to dry the IPA on the surface of the workpiece. The object was then put in a stream of nitrogen or argon while being flashed using UV light for 10~15 minutes.

**Scanning Electron Microscopy (SEM)** SEM was applied to visualize the structure of the 3D-printed sample surfaces with emphasis on feature size and orientation. The investigations were carried out using two different field emission gun (FEG) scanning electron microscopes: a). A Helios Hydra 5 UX FEG-SEM & Plasma FIB (focused ion beam) dual beam system, provided by Thermo Fisher, for closely studying small structural details with a short working distance (5-10mm). b). A NanoNova600 FEG SEM, provided by FEI (now Thermo Fisher), for overview imaging of wider sample areas with as large as possible working distance (ca. 40mm). Secondary electron (SE) imaging, using Everhart Thornley (ETD) SE-detectors in both instruments, was applied to display the topography of the sample surface. To avoid electrostatic charge (imbalance between negative and positive charge on the images sample spot), samples were coated with a ~20 nm carbon layer and in general as low as possible acceleration voltage was applied (not higher than 5kV). Samples were imaged in several tilted angles for a better 3-dimensional overview.

**Reactive Transport Simulation** Numerical simulation based on a previously developed DC-TVP model[19] was used to gain general insights into the system dynamics and to facilitate experimental trial-and-error. In a simulation, the 3D curing volume was discretized using the same voxelization scheme in sinogram computation and the reactive transport of oxygen (A), lophyl radical (B) and *o*-Cl-HABI (C) was analyzed by solving

$$\frac{dc}{dt} = \nabla^2 c + Da \quad . \tag{12}$$

Using oxygen as an example, *c* is the dimensionless inhibitor concentration

$$c = \frac{[A]_0 - [A]}{[A]_0}. \qquad (13)$$

[A] and [A]₀ are the concentration and the initial concentration of oxygen in the polymer precursors, respectively (mol/L). Here, *Da* is the Damköhler number:

$$Da = l^2 \frac{k_0 \int_\lambda \alpha(\lambda) I(\lambda) d\lambda}{[A]_0 D_A}, \qquad (14)$$

in which $l$ is voxel size (m), $D_A$ is the diffusivity of inhibitor (m²/s), $\alpha$ is the absorption coefficient and *I* the irradiance (mW) of wavelength $\lambda$ (nm), $k_0$ is the zeroth order rate constant that relates inhibitor consumption to irradiance. *Da* evolves spatiotemporally. The Fickian fluxes $N_A$ for each voxel were computed at the six inter-voxel surfaces and photochemical reactions were treated as a zeroth order sink for oxygen, first order sink for *o*-Cl-HABI and second order sink for lophyl radicals, in accordance with Equations 5-7. It was necessary to treat the photolysis of *o*-Cl-HABI as a first order reaction to account for the limited availability of lophyl radicals. Concentration was assumed uniform inside each voxel. Such simulations were used to helped optimize exposure time for DC-TVP experimentally, and was not part of the projection optimization. Movies S1 and S9-11 were results of such numerical simulations. A MATLAB implementation of the reactive-transport model is supplemented as Script S4.

**Acknowledgments:** We thank Ecem Badruk, Christof Hieger, Arturo Bianchetti and IN-VISION Technologies for providing a TwoWave light engine for testing; Esben Thormann for access to optical microscope; Josefine F. Lønholdt, Jane Pedersen, Yan Wei, Stefan Bruns and Christina Schmidleithner for helpful discussions and resource acquisition; Mariusz Kubus for assistance in FT-IR measurement; Peter R. Stubbe for guidance in mechanical testing; Ishaq Khaliqdad for TVP prototyping; Christina B. Nielsen for assistance in chemical analysis and management. We are particularly grateful to Dr. Kai Neufeld and DTU Nanolab for SEM analysis, and to the 3D Imaging Center at the Technical University of Denmark for access to X-ray computed tomography, We thank National Gallery of Denmark (SMK) and the Thingiverse community for creating high quality STL files for 3D printing research.

**Author contributions:**

Conceptualization: YY

Methodology: BW, HS, YY

Investigation – experimental: BW, WS, HS, SKH, JPCN, TDVC, AAS

Investigation – modelling: YY

Funding acquisition: JRF, KA, AI, YY

Project administration: YY

Supervision: JRF, HOS, KA, AI, YY

Writing – original draft: BW, HS, YY

Writing – review & editing: All authors

**Competing interests:** WS and YY are shareholders of PERFI Technologies ApS (DK44698978). The remaining authors declare no competing interests.

**Data availability:** The authors declare that all data supporting the findings of this study are available within the paper and its supplementary information files.

**Code availability:** MATLAB realization of mathematical algorithms central to the conclusions of this study are provided in the supplementary information files.

**Supplementary Information**

17 supplementary figures (Figs S1 – S17)

11 supplementary movies (Mov S1 – S11)

4 supplementary MATLAB scripts (Scripts S1 – S4)

1 supplementary data spreadsheet (Supplementary Data 01.xlsx)